\documentclass[preprint2,longabstract]{aastex}

\renewcommand{\b}[1]{\mbox{\boldmath $#1$}}

\def\cal#1{{\cal #1}}

\def\m@th{\mathsurround=0pt}
\def\n@space{\nulldelimiterspace=0pt \m@th}
\def\biggg#1{{\mbox{$\left#1\vbox to 20.5pt{}\right.\n@space$}}}

\def\beginenum{\begin{enumerate}}
\def\endenum{\end{enumerate}}
\def\bitem{\begin{itemize}}
\def\eitem{\end{itemize}}
\def\bray{\begin{array}}
\def\eray{\end{array}}
\def\begindoc{\begin{document}}
\def\enddoc{\end{document}}
\def\bq{\begin{equation}}
\def\eq{\end{equation}}
\def\bqy{\begin{eqnarray}}
\def\eqy{\end{eqnarray}}
\def\bqyn{\begin{eqnarray*}}
\def\eqyn{\end{eqnarray*}}
\def\bc{\begin{center}}
\def\ec{\end{center}}
\def\bfll{\begin{flushleft}}
\def\efll{\end{flushleft}}
\def\bflr{\begin{flushright}}
\def\eflr{\end{flushright}}
\usepackage{spr-astr-addons}    %% mimicing ASTR journal style
\usepackage{color}

\RequirePackage{color}
\def\imagei{\centerline{\color[gray]{.75}\rule{\hsize}{4pc}}}%
\def\imageii{\centerline{\color[gray]{.75}\rule{4pc}{4pc}}}%

\newcommand{\vdag}{(v)^\dagger}
\newcommand{\emaila}{authors@email.com}

\begin{document}
%% Article title
%
%\title{Unified Reverse Dynamo/Dynamo mechanism in 2-Temperature relativistic
%electron-ion plasmas of astrophysical objects}

\title{Macro-scale fast flow and magnetic field generation in
2-temperature relativistic electron-ion plasmas of astrophysical objects}
%due to Unified Reverse Dynamo/Dynamo mechanism}

%% Running heads
\shorttitle{<Unified RD/D in 2-temperature relativistic plasmas>}
\shortauthors{< Kotorashvili \& Shatashvili >}

\author{ K. Kotorashvili\altaffilmark{1}}
\altaffiltext{1}{Department of Physics, Faculty of Exact \&
Natural Sciences, Javakhishvili Tbilisi State University,
Tbilisi 0179, Georgia}
\author{ N.L. Shatashvili\altaffilmark{1,2,*}}
%\email{nana.shatashvili@tsu.ge}
\altaffiltext{1}{Department of Physics, Faculty of Exact \&
Natural Sciences, Javakhishvili Tbilisi State University,
Tbilisi 0179, Georgia}
\altaffiltext{2}{Andronikashvili Institute
of Physics, TSU, Tbilisi 0177, Georgia}
\altaffiltext{*}{E-mail: nana.shatashvili@tsu.ge}

\begin{abstract}
We have shown the simultaneous generation of macro-scale fast flows
and strong magnetic fields in the 2-temperature relativistic electron-ion
plasmas of astrophysical objects due to Unified Reverse Dynamo/Dynamo mechanism.
The resulting dynamical magnetic field amplification and/or flow acceleration
is directly proportional to the initial turbulent kinetic/magnetic (magnetic)
energy; the process is very sensitive to relativistically hot
electron-ion fraction temperature and magneto-fluid coupling.
It is shown, that for realistic physical parameters of
White Dwarfs accreting hot astrophysical flow
/ Binary systems there always exists such a real solution of dispersion relation
for which the formation of dispersive strong super-Alfv\'enic macro-scale flow/outflow
with Alfv\'en Mach number $> 10^6$ and/or generation of super-strong magnetic fields
is guaranteed.
%as observed in accreting astrophysical objects.

\end{abstract}

%\startpage{1}
%\endpage{1}

\maketitle

%% Keywords

\section{Introduction}

Multi-temperature composite systems are often met in various
astrophysical settings, e.g. a highly degenerate
White Dwarf (WD) plasma co-existing with a classical hot accreting
astrophysical flow. WDs - interesting representatives of compact astrophysical
objects - comprise up to $98\%$ of the end state of all stars
\citep{winget,Compact-WD,kulebi,kepler}. 
It is considered that
accreting WDs (AWD) are featuring global magnetic structures
with field strengths ($1 - 1000$)\,MG \citep{White,hmfwd};\\
\citep{Kawka}. Isolated WDs can be separated into two groups
- high field magnetic WDs (HFMWDs) with magnetic fields stronger
than $10^6$\,G and the rest with lower magnetic fields tipically
$<10^5$\,G. About 10\% of isolated WDs are HFMWDs
\citep{DAWD,Kawka,Ferrario-3}. Recent studies point toward
a binary origin of HFMWDs (see e.g. \citep{Garcia}
and references therein).

Many stars are born in the binary systems going through one or more
phases of the mass-exchange \citep{winget,DAZ};\\
\citep{tremblay,mukai}. Numerous observed accreting
WDs are often surrounded by an accretion gas of companion
star / disk \citep{Begelman,mukai}. Cataclismic Variables (CVs)
and Symbiotics are representing the accreting white dwarf
binaries (AWBs). A WD with a close companion that is overflowing
its Roche lobe is a cataclysmic variable (CV) \citep{Warner,DAWD}.
Depending on the magnetic field, there are two classes of CVs:
nonmagnetic CVs characterised by their eruptive behavior
\citep{Warner,Balman,mukai}, with weak or nonexistent
magnetic fields ($< 0.01MG$) and the magnetic CVs (MCVs),
divided into two sub-classes as Polars (with strong magnetic
fields in the range of $(20 - 230)$\,MG, which cause the accretion
flow to directly channel onto the magnetic poles of the WD
inhibiting the formation of an accretion disk) and Intermediate
Polars, which have a weaker field strength of (1-20)\,MG,
\citep{Warner,Mouchet,mukai}. Among the CVs about $25\%$
\citep{Ferrario-2} have WDs that are very magnetic \citep{Balman}.

It was proposed recently that WD magnetic fields have
a fossil origin \citep{Fossil} originally suggested for Sun \citep{Couling}.
An alternative suggestion is that they could be generated
by a dynamo process in the star's convective core \citep{ferrario} or/and during
the star's evolution common envelope phase by accreting disk-dynamo \citep{Disk-DB}.
However, fossil origin idea has difficulty explaining high field strengths
and the observed lack of a correlation with rotation as well
as the absence of field configurations stable enough to survive
in a star over its lifetime \citep{Fossil}.

According to \citep{Ferrario-3} the fact that WDs with
a surface magnetic field over 3\,MG has not been found in
a detached binary system suggests that all such highly
magnetic WDs have a binary origin, which relies on a
magnetic dynamo (D) during the common envelope (CE) phase
of binary evolution. If the two stars merge the end product
is a single HFMWD that may later evolve into the MCVs \citep{Ferrario-4}.

In addition to the magnetic field it is expected, that large-scale
outflows met in various astrophysical setting being the
collimated long-lived structures related to accreting disks
surrounding the compact objects [see e.g. \citep{Begelman}
and references therein] are also playing an important role
in stellar evolution, including late stages of their lives
and their final fates. AWBs are important laboratories for
accretion and outflow physics. In CVs (nova-likes) the outflow
velocities can vary within 200-5000 km/s \citep{AD_modeling,kafka,Diaz}.

For a typical cold magnetic WD with degenerate electron densities
$\sim (10^{25}-10^{29})cm^{-3}$ and magnetic fields
$\sim (10^{5}-10^{9})$\,G, temperatures are $\sim (40000-6000)$\,K
\citep{DAZ,hollands}. At very high densities, particle Fermi
Energy can become relativistic and degeneracy pressure (${\mu }_0$)
may dominate thermal pressure ($\ {\beta }_0$); for these parameters,
Alfv\'{e}n velocity $V_A\sim ({10}^4 - {10}^6) $ cm/s ,
yielding plasma $\ {\beta }_0\sim ({10}^6-\ {10}^0) $, ${\mu }_
0\sim ({10}^{10}-\ {10}^6) $ leading to ${\mu }_0 \gg \ {\beta }_0$,
so the plasma may be treated as cold, even
with temperature $\sim 10^9 K$ \citep{BSM_deg}.

The consequences of degeneracy in multi-component plasma were
studied recently in terms of multi-scale behavior \citep{BSM_deg,SMB_multi};\\
\citep{BS-flow} to investigate effects of degeneracy e.g.,
on the dynamics of the star collapse that seems to be sensitive
to outer layers/atmosphere composition, structure and their
conditions. These studies have shown the generation/amplification
of fast macro-scale plasma flows in the degenerate two-fluid
astrophysical systems, for which the simplest relaxed states are
described by double Beltrami condition, which states that the generalized
flow is aligned with its vorticity augmented by the Bernoulli condition.

Shatashvili et al (2019) have shown that in a quasi neutral plasma
of a bulk degenerate electrons contaminated by small fraction of
a non-degenerate highly relativistic hot electron component
can induce a new scale (for structure formation) to a system
consisting of an ion-degenerate electron plasma. Determined by
concrete parameters of the system, new macro-scale lengths
(much larger than short intrinsic scale lengths [skin depths]
and generally much shorter than system size) open new pathways
for energy transformations. It is expected that this combination
of plasmas will also pertain during the relativistic jet formation from
accretion-induced collapsing WDs to Black Holes \citep{Begelman,Kryvdyk,JetsWD}.

The main goal of the paper is to find the possible mechanism for the origin
and evolution of large-scale magnetic/velocity fields in multi-component
compact astrophysical objects and their vicinity; more precisely
we will examine the possible role played by the Unified Reverse Dynamo (RD)/Dynamo
in explaining the exitance of macro-scale flow and magnetic field
in a highly degenerate ($d$) WD plasma co-existing with
a classical hot ($h$) accreting astrophysical flow. We will investigate the
role of degeneracy effects and explore the new physics originating from
the contamination of the ($h$) component in the formation of macro-scale magnetic
and velocity fields to uncover the effects of magneto-fluid couplings of accreting
stars since the dynamical evolution of their convective envelopes may define
the final structure of their interior as well as of atmospheres.

\section{Model Equations for Unified Reverse Dynamo / Dynamo mechanism for %AWDs
2-Temperature Relativistic e-i plasmas of astrophysical objects}

In our recent paper \citep{RD_deg} from an analysis of the degenerate
two-fluid (electron-ion or electron-positron) system,
we have extracted the Unified RD/D mechanism \citep{msms,lingam}
- the amplification / generation of fast macro-scale plasma
flows in astrophysical systems with initial turbulent (micro-scale)
magnetic/velocity fields. It was shown that the genarated/acceletared
locally super-Alfvenic flows are extremely fast with Alfv\'{e}n
Mach number $>10^3 $ as observed in different astrophysical outflows.
This process is simultaneous with and complementary to the
micro-scale unified dynamo - generation of macro-scale flows
is an essential consequence of magneto-fluid coupling; generation
of macro-scale fast flows and magnetic fields are simultaneous,
they grow proportionately.

In this section we will be studying a quasi neutral plasma
of an immobile classical ion component ($i$), and two relativistic
electron species - the bulk $d$ electron gas with a density
$N_{0d}$ and a small contamination of $h$ electrons with density
$N_{0h}$. In our simplified model the gravity (Newtonian) effects
are ignored when solving the Dynamo/Reverse Dynamo problem to show
the tendency of such complex system to generate large-scale
Magnetic/Velocity fields locally; hence, the density variations are ignored,
such effects are important at the distances where the catastrophic
transformation of energy takes place \citep{mnsy,BS-flow};
and the wave-coupling phenomena (see e.g. \citep{SMB_ElSound})
are beyond the scope of present study; note, that
for a disk-jet structure formation both the gravity and rotation effects
were studied in \citep{SY-DJ,yso}. This model can be applied
for the description of the system of dense/degenarte
WD's outer layer that
accretes a classical hot astrophysical flow. Also, based on
the observations, we assumed that hot electron fluid fraction is
small and in such three component plasma ion fluid velocities are
much smaller than those for lighter electron ($d,h$) fluids
[$\textbf{V}_i << \textbf{V}_d,\textbf{V}_h$].
Thus, ion dynamics is neglected  \citep{relaxed}. Then, following
Shatashvili et al. (2019) we write for quasi neutrality condition
\begin{equation}
N_{0d} + N_{0h} = N_{0i} \ \ \Longrightarrow \ \
\frac{N_{0i}}{N_{0d}} = 1+\alpha, \quad   \alpha \equiv \frac{N_{0h}}{N_{0d}} \  ,
\label{2TRD-1}
\end{equation}
where \ $\alpha \ll 1$ \ labels the ratio of hot electron fraction
to the degenerate electrons. Notice, that flow effects were found
crucial in creating the structural richness, in the heating/cooling
processes, in Generalized Dynamo theory and outflow formation in
astrophysical environments \citep{mmns-1,mnsy,msms,lingam};
\citep{RD_deg} and since in our system there are two symmetry-breaking
mechanisms: 1) one is due to different effective inertias for the
$d $ and $h$ electrons, and 2) the other is from the  small $h$
contamination added to the bulk $d$ electrons ($\alpha \neq 0 ,
{\bf V}_h \neq 0$), both of them are responsible for creating
a net ``current''. The structure formation mechanism
originates, for instance, in the effective inertia
difference (see e.g. \citep{BM-94,structuresPI-2} and references
therein). Asymmetry between the plasma constituents increases
the number of conserved helicities, and eventually translates into
a higher index equilibrium Beltrami state \citep{Multi-B,SMB_multi};\\
\citep{SMB_2T} and creation of electron-sound \citep{SMB_ElSound}
in such complex relativistic environment. We will show below that
even in the case of immobile ions when the rotation
effects are ignored (important for pulsars/pulsar binaries \citep{PulsarBinary}
and not the case of present study) these asymmetries strongly define
the dynamics of Unified Dynamo/Reverse dynamo action and lead
to very interesting and new scenarios for fast super-Alf\'enic
flow/outflow formation or/and strong magnetic field
formation in the vicinity of WDs that accrete hot
astrophysical flow.

We remind the reader that the effective mass factor $G_{d}(n_{d})=
\sqrt{1+(\frac{n_{d}}{n_{c}})^\frac{2}{3}}$  for degenerate
electron plasma originates from degeneracy rather then kinematics
and is fully determined by the plasma rest frame density $n_d=N_d/\gamma_d$
(see \citep{BSM_deg} and references the in) for arbitrary $n_{d}/{n_{c}}$
(with $n_c=5.9\cdot {10}^{29}{cm}^{-3}$ being the critical number-density).
While for relativistically hot electron plasma the effective mass factor
$G_{h}=\frac{5}{2}\frac{T_{e}}{m_ec^2} +\frac{3}{2}\sqrt{\left(\frac{T_{e}}
{m_ec^2}\right)^2+\frac{4}{9}} $ \citep{mignone,Ryu}
is determined by the relativistic electron temperature $T_e$.
As mentioned above below we consider the analytically tractable
constant-density system to study the Unified Dynamo/RD phenomena
in our complex 2-temperature relativistic system, assuming
quasi-neutrality $\varphi\equiv 0,$ and for simplicity we put
$ \ \gamma_d \sim \gamma_h\sim 1 $ leading to $ G_d=const\equiv G_0(n_{0d}) ,
\ G_h=const\equiv H_0(T_{e0}) $ \citep{SMB_2T}.

The electron dynamics for both components can be described
by the appropriate relativistic fluid equations (see e.g. \citep{BSM_deg}
and references therein): the continuity and the equation
of motion \citep{SMB_2T} that can be cast into an ideal vortex dynamics
in terms of the generalized (canonical) vorticities. Our model
pertains only for homentropic plasmas; for special astrophysical
conditions, canonical vorticity would have a quantum-mechanical
part [see \citep{SpinVort,Felipe} for spinning plasmas], and even a general
relativistic component [see e.g. \citep{AccrVort} for
Black Hole accretion disks] in addition to the electromagnetic,
kinetic and thermal contributions; for most applications
these corrections are negligibly small. Moreover, even for the
highly relativistic case when the space-time metric is almost
Minkowskian and terms of order $O(c^{-4})$ or higher are neglected,
it is possible to write the (linearized) equations of general
relativity (GR) in a form that is almost identical to that of
the Maxwell equations of ordinary electromagnetism
(see e.g. \citep{Manfredi} and references therein).
Then, in the dimensionless form these equations for 2-temperature
relativistic plasma consisted of bulk degenerate e-i plasma
contaminated by relativistically hot electron ion plasma in the
limit of immobile ions can be reduced to (where ${\bf b}$ and
${\bf V}_{d(h)}$ are the dimensionless magnetic and velocity
fields, respectively):
\[
\bigg(G_0+\frac{H_0}{\alpha}\bigg)\frac{\partial {\bf b}}{\partial t} \
+ \ G_0\frac{H_0}{\alpha}\frac{\partial }{\partial t} \nabla \times
\nabla \times {\bf b} \ =
\]
\[
= \frac{(H_0-G_0)}{\alpha}\nabla\times {\bf {V}}_d \times {\bf b}
\ + \ \frac{G_0}{\alpha}\nabla\times({\bf b}\times (\nabla\times {\bf b}) \ -
\]
\[
\quad - \ \frac{G_0 H_0}{\alpha^2} \nabla \times ({\bf {V}}_d \times
\nabla \times {\bf V}_d \ + \ \textbf V_d \times \nabla \times
\nabla \times {\bf b} ) \ -
\]
\[
\qquad - \ \frac{G_0 H_0}{\alpha^2}\nabla\times\left(
(\nabla \times {\bf b}) \times \nabla \times{\bf {V}}_d \right) \ -
\]
\begin{equation}
\qquad - \ \frac{G_0 H_0}{\alpha^2}\nabla\times\left(
(\nabla \times {\bf b}) \times \nabla \times (\nabla \times {\bf b})\right) \ ,
\label{2TRD-2}
\end{equation}
\[
\bigg(G_0+\frac{H_0}{\alpha}\bigg)\frac{\partial {\bf V}_d }{\partial t} \ +
\ \frac{H_0}{\alpha}\frac{\partial }{\partial t}\nabla \times {\bf b}
\ = - \ \frac{1}{\alpha}{\bf V}_d \times {\bf b} \ +
\]
\[
\qquad + \ \frac{1}{\alpha} {\bf b}\times (\nabla\times {\bf b)}
- \ \bigg(\frac{H_0}{\alpha^2}-G_0\bigg)\textbf V_d \times
\nabla \times {\bf V}_d \ -
\]
\begin{equation}
\qquad - \ \frac{H_0}{\alpha^2}\,{\bf V}_d \times \nabla \times
(\nabla \times {\bf b}) \ +
\label{2TRD-3}
\end{equation}
\[
+ \ \frac{H_0}{\alpha^2}\left((\nabla \times {\bf b}) \times
\nabla \times {\bf V}_d+(\nabla \times {\bf b})
\times \nabla \times \nabla \times {\bf b}\right) ,
\]
that can be closed by Ampere's law
\begin{equation}
\nabla\times {\bf b} \ = \ - \ {\bf V}_d \ - \ \alpha \,{\bf V}_h  \ ,
\label{2TRD-4}
\end{equation}
where density is normalized to $N_{0d}$ (corresponding rest-frame
density is $n_{0d}$); the magnetic field is normalized to
some ambient measure $|{\bf B}_0|$; hot electron gas
temperature is normalized to $m_ec^2$; all velocities
are measured in terms of the corresponding Alfv\'en speed $V_A= V_{Ad}
={B_0}/{\sqrt{4\pi n_{0d}m_eG_{0d}}}$ all lengths [times] are
normalized to the ''effective'' degenerate electron skin depth
$\lambda_{\rm{eff}}^d[\lambda_{\rm{eff}}^d/V_A]$ with $
\lambda_{\rm{eff}}^d={c}/{\omega_{pe}^d}
=c\sqrt{{m_e G_{0d}}/{4\pi n_{0d}e^2}} $.

Following the standard procedure Mahajan et al. (2005) let's
assume that our total fields are composed of some ambient
seed fields and fluctuations about them taking into account
relativistic effects both for bulk degenerate
electron-fluid and hot electron fraction:
\begin{equation}
{\bf b} =  {\bf b}_0 + {\bf B} + \tilde{\bf b} ,
\ \ \    {\bf V}_{d(h)} = {\bf v}_{0d(0h)} + {\bf U}_{d(h)}
+ \tilde{\bf v}_{d(h)} \ ,
\label{2TRD-5}
\end{equation}
where ${\bf b}_0$ and ${\bf v}_{0d(0h)}$  are the equilibrium fields,
${\bf B}$ , ${\bf U}_{d(h)}$ are the macroscopic fluctuations and
$\tilde{\bf b}$, $\tilde{\bf v}_{d(h)}$ are the microscopic
fluctuations, respectively. We remind that the energy reservoir comes
from the background fields, which have both macroscopic and microscopic
components feeding the macro- and micro-scale fluctuations of fields.
To describe background fields, for analytical work,
we choose a special class of equilibrium Beltrami-Bernoulli (BB)
solutions to the equations (\ref{2TRD-2}) and
(\ref{2TRD-3}) (see e.g. \citep{msms,lingam,RD_deg}) for both degenerate $d$
and hot $h$ electrons well studied in \citep{SMB_2T}
that reduce to:
\begin{equation}
\textbf b_0-G_0 \nabla \times {\bf v}_{0d}=a_d {\bf v}_{0d} ,
\ \   {\bf b}_0-{\bf H}_0 \nabla \times {\bf v}_{0h}
=\alpha a_h {\bf v}_{0h}
\label{2TRD-6}
\end{equation}
which, following the straightforward algebra, reduces to
a Triple Beltrami (TB) equation for ${\bf b}_0$:
\[
 G_0\frac{H_0}{\alpha} \nabla \times \nabla \times \nabla
 \times {\bf b}_0 + \bigg(a_d\frac{H_0}{\alpha} + a_hG_0\bigg)
 \nabla \times \nabla \times {\bf b}_0
 \]
 \begin{equation}
+\bigg(G_0+\frac{H_0}{\alpha}+a_da_h \bigg)\nabla\times {\bf b}_0
+\bigg(a_d+a_h\bigg){\bf b}_0=0 ,
\label{2TRD-7}
\end{equation}
and relations for ${\bf v}_{0d}$ and ${\bf v}_{0h}$
($\eta \equiv a_d\,\frac{H_0}{\alpha} - a_hG_0$):
\[
{\bf v}_{0d} \ = \ \eta^{-1}\,
\left(G_0\frac{H_0}{\alpha}\nabla\times\nabla\times\nabla\times\textbf b_0
\right)\ +
\]
\begin{equation}
\qquad + \ \eta^{-1}\,\left[a_hG_0 \nabla \times {\bf b}_0 \
+ \ \left(G_0 + \frac{H_0}{\alpha}\right)\,{\bf b}_0\right] \ ,
\label{2TRD-8}
\end{equation}
\[
{\bf v}_{0h} \ = \ - \ \frac{\eta^{-1}}{\alpha}\,
\left(G_0\frac{H_0}{\alpha}\nabla\times\nabla\times\nabla\times {\bf b}_0
\right)\ -
\]
\begin{equation}
\qquad - \ \frac{\eta^{-1}}{\alpha}\,\left[a_d\frac{H_0}{\alpha} \nabla
\times {\bf b}_0 \ + \ \left(G_0 + \frac{H_0}{\alpha}\right)\,{\bf b}_0\right] ,
\label{2TRD-9}
\end{equation}
where $\nabla\cdot \textbf b_0 = 0, \ \ \nabla\cdot \textbf v_{0d(0h)}=0 $
are automatically satisfied and $a_{d(h)}$ are dimensionless constants
related to two invariants:  \ $h_{d(h)} = \int\,{d^3x}\,{({\bf A}\pm G_0(H_0)
{\bf v}_{0d(0h)})}\,\cdot $ \\
$({\bf b}_0\pm G_0(H_0)\nabla \times {\bf v}_{0d(0h)})$
- the generalized helicities of $d(h)$ relativistic electron fluids.
Here the generalized vorticities for both fluids have now both
magnetic and kinetic parts due to the corresponding relativistic effects.
We need to choose these constants (that together with the defining
parameters $G_0 , H_0 , \alpha$ fully determine the equilibrium system)
so that the scales [solutions of the equation \ $ G_0\frac{H_0}{\alpha}\mu^3
+\big(a_d\frac{H_0}{\alpha}+a_hG_0\big)\mu^2
+\big(G_0+\frac{H_0}{\alpha}+a_da_h \big)\mu+\big(a_d+a_h\big) =0$ ]
\ are vastly separated. Following the similar procedure as in
\citep{msms,lingam,RD_deg}, assuming that main energy reservoir is in
micro-scale equilibrium component and $|\tilde{\bf b}|\ll |{\bf b_0}| ,
\ |\tilde{\bf v}_{d(h)} |\ll |{\bf v_{0d(0h)}}| $   we find that
the equilibrium velocity and magnetic fields are related as
\begin{equation}
{\bf v}_{0d(h)}= \chi_{d(h)}{\bf b}_0 \ ,
\label{2TRD-10}
\end{equation}
where $\chi_d$ and $\chi_h$ are given by
\[
\chi_d\equiv \eta^{-1}\left[G_0\frac{H_0 }
{\alpha}\lambda^2+a_hG_0\lambda  + \left(G_0
+ \frac{H_0}{\alpha}\right)\right] \ ,
\]
\begin{equation}
\chi_h\equiv-\frac{\eta^{-1}}{\alpha}\left[G_0\frac{H_0 }{\alpha}\lambda^2
+a_d\frac{H_0}{\alpha}\lambda + \left(G_0 + \frac{H_0}{\alpha}\right)\right].
\label{2TRD-11}
\end{equation}

\bigskip

The straightforward algebra gives the following equations
for micro-scale magnetic and degenerate fluid velocity fields
(the derivations for relativistically hot fluid component relations
see the Appendinx-A):
\begin{equation}
m\frac{\partial \tilde{\bf b}}{\partial t} \
+ \ m_1\frac{\partial }{\partial t} \nabla \times \nabla \times
\tilde{\bf b}\ = \ ({\bf Q}_d \cdot \nabla){\bf b}_0 \ ,
\label{2TRD-12}
\end{equation}
\begin{equation}
m\frac{\partial \tilde{\bf v}_{d}}{\partial t}
\ + q_1\frac{\partial }{\partial t}\nabla\times\tilde{\bf b}\
= \ ({\bf S}_d \cdot \nabla ){\bf b}_0 \ ,
\label{2TRD-13}
\end{equation}
where ${\bf Q}_d$ and ${\bf S}_d $ are functions of ${\bf U}_d$ and
${\bf B}$, as well as plasma-system parameters. Using (\ref{2TRD-12}),
(\ref{2TRD-13}) we obtain the final equations for macro-scale ${\bf B}$
and ${\bf U}_d$:
\begin{equation}
m_1\,{\nabla \times \nabla \times \ddot{\bf B}} + m \,{\ddot{\bf B}}
= r_d\,\nabla\times {\bf B} + p_d\,\nabla\times {\bf U}_d .
\label{2TRD-14}
\end{equation}
\begin{equation}
q_1\,{\nabla \times \ddot{\bf B}} + m\,{\ddot{\bf U}}_d
= q_d\,\nabla \times {\bf B} + s_d\, \nabla \times {\bf U}_d ,
\label{2TRD-15}
\end{equation}
where
\[
m_1=\frac{G_0H_0}{\alpha}\bigg(G_0 + \frac{H_0}{\alpha}\bigg) \ , \ \ \
m=\bigg(G_0+\frac{H_0}{\alpha}\bigg)^2 \ ,
\]
\begin{equation}
q_1=\frac{H_0}{\alpha}\bigg(G_0+\frac{H_0}{\alpha}\bigg) \ ,
\label{2TRD-16}
\end{equation}
and the constants
\[
r_d = \frac{\lambda b^2_0}{3}\bigg[\frac{H_0-G_0}{\alpha^2}
-\frac{G_0H_0}{\alpha^3}\lambda(\chi_d +\lambda) \ +
\]
\[
\qquad + \ \bigg(\frac{G_0}{\alpha}(\chi_d+\lambda)-\frac{H_0}{\alpha}
\chi_d\bigg)^2\bigg] \ ,
\]
\[
p_d = - \frac{\lambda b^2_0}{3}\bigg[\frac{H_0-G_0}{\alpha}
-\frac{G_0H_0}{\alpha^2}\lambda(\chi_d+\lambda)\bigg]
\bigg(\frac{H_0-G_0}{\alpha^2\lambda} \ -
\]
\[
\qquad - \ \frac{G_0H_0}{\alpha^3}(\chi_d+\lambda)
+\frac{H_0}{\alpha^2}(\chi_d+\lambda)
-\chi_dG_0 \ -
\]
\[
\qquad - \ \frac{G_0}{\alpha}(\chi_d+\lambda)-\frac{H_0}{\alpha}\chi_d\bigg) \ ,
\]
\[
q_d = - \frac{\lambda b^2_0}{6}\bigg[\bigg(\frac{G_0}{\alpha}
(\chi_d+\lambda)-\frac{H_0}{\alpha}\chi_d\bigg)\cdot
\]
\[
\qquad \cdot \bigg(\frac{H_0^2}{\alpha^2q_1}\chi_d\lambda(G_0+1)
+\frac{G_0H_0}{\alpha^2q_1}
+\frac{H_0}{\alpha^2}\lambda(\chi_d+\lambda)\bigg) \ -
\]
\[
\qquad - \ \frac{H_0}{\alpha^3}(\chi_d+\lambda)
+\frac{G_0}{\alpha}\chi_d\bigg],
\]
\[
s_d = - \ \frac{\lambda b^2_0}{6}\bigg[\bigg(\frac{H_0}
{\alpha^3\lambda}(\chi_d+\lambda)
- \ \frac{G_0}{\alpha\lambda}\chi_d \ +
\]
\[
\qquad + \ \frac{H_0}{\alpha^2}(\chi_d+\lambda)-\chi_dG_0\bigg)\ +
\]
\[
\qquad + \ \bigg(\frac{(H_0-G_0)}{\alpha}
-\frac{G_0H_0}{\alpha^2}\lambda(\chi_d+\lambda)\bigg) \cdot
\]
\begin{equation}
\qquad \cdot \bigg(\frac{H_0^2}{\alpha^2q_1}\chi_d\lambda(G_0+1)
+\frac{G_0H_0}{\alpha^2q_1}+\frac{H_0}{\alpha^2}\lambda
(\chi_d+\lambda)\bigg)\bigg] \ ,
\label{2TRD-17}
\end{equation}
where $b_0^2(v_0^2)$ measures the ambient micro-scale (turbulent)
magnetic(kinetic) energy and all the coefficients are functions
of background system defining parameters $G_0, H_0, \alpha ,
\lambda, a_d, a_h$ (hence, determined by energies, effective
masses, fraction coefficient and helicities).

Performing a Fourier analysis we obtain the following dispersion
relation (DR):
\[
\omega_d^8 m^{\prime \,2} m^2 \ - \ \omega_d^4k^2\Big(2m^\prime m p_d
q^\prime_d+m^2r_d^2+m^{\prime \,2}s_d^2\Big) \ +
\]
\begin{equation}
\qquad + \ (p_dq^\prime_d-r_ds_d)^2k^4 = 0 \ ,
\label{2TRD-18}
\end{equation}
$$
{\rm with}  \qquad \qquad m^\prime=(m+m_1k^2), \ q^\prime_d=(q_d+q_1\omega^2)
$$
and, finally, using the same procedure for hot relativistic
fluid (presented in Appendix-A) we derive the relations for
all macro-scale fluctuations of vector-fields (with
corresponding DRs (\ref{2TRD-18}) and (\ref{2TRD-A7})):
\begin{equation}
\hspace{3.5cm} {\bf U}_{d(h)} \ =
\label{2TRD-20}
\end{equation}
\[
= \ \frac{q^\prime_{d(h)}\Big[\omega_{d(h)}^4 m^\prime m
-(p_{d(h)}q^\prime_{d(h)}-r_{d(h)}s_{d(h)})k^2\Big]}
{\omega_{d(h)}^4 m^2r_{d(h)}+s_{d(h)}[\,p_{d(h)}q^\prime_{d(h)}
-r_{d(h)}s_{d(h)}\,]k^2} \ {\bf B} \ ,
\]
\noindent Equations (\ref{2TRD-20}) are the manifestation of
Unified Dynamo/Reverse Dynamo - a simultaneous generation of
large-scale magnetic field and flow from ambient short-scale
turbulent energy (magnetic or/and kinetic) -
a manifestation of magneto-fluid coupling. Interestingly,
relations between fields have the same form for degenerate
and hot fluids but we shall remember, that defining coefficients
[$r_{d(h)}, p_{d(h)}, q_{d(h)}, s_{d(h)}$] are different
for these fluids as well as the solutions of corresponding
dispersion relations are different.
Thus, the larger the ambient system turbulent energy is, the stronger
the flow acceleration will be; for more details the reader may consult
with \citep{msms,lingam,RD_deg}). \\
We also observe that to
leading order, due to hot relativistic fluid contamination,
evolution of ${\bf B}$ does require knowledge of ${\bf U}_d$
and vice versa (compare with degenerate e-i case \citep{RD_deg},
when knowledge of ${\bf B}$ does not require knowledge of ${\bf U}_d$);
note, that now similar conclusion is valid for ${\bf U}_h$;
hence, magneto-fluid coupling process is much richer now leading
to various scenarios for magnetic field generation or/and flow
acceleration both for degenerate and hot components. In the next
section we will show how the choice of system parameters define
final fate of the system. From various scenarios
we chose 2 limiting simplest cases for illustration
(interesting for WDs accreting hot astrophysical flow): \\
\noindent (i) $a_d = -a_h\equiv a$  and Beltrami index goes from
triple to double. For WDs with $n_0 \equiv 10^{25}-10^{29} cm^{-3}$
leading to $G_0\geq 1.5$ and for $H_0 > 1 , \ \alpha \ll 1 $ \
we have \ $\frac{H_0}{\alpha} \gg G_0$.
When choosing the inverse micro-scale to be \
$\lambda = -\frac{\alpha}{2G_0H_0}\left[a\Big(\frac{H_0}{\alpha}
-G_0\Big)+\Big(\frac{H_0}{\alpha}+G_0\Big)
\sqrt{a^2 -\frac{4G_0H_0}{H_0+\alpha G_0}}\right]$  \\
we find that for \  $a > G_0+1$ \ the ambient energy is mostly
kinetic while for \ $2\sqrt{G_0} < a < G_0 + 1$ \ the
ambient energy is mostly magnetic. \\
\noindent (ii) $a_d = \frac{1}{a_h}$ , for which numerically
solving the the relevant equation to find scales we could
find the regimes to determine the dominance of magnetic/kinetic
energies in the ambient energy of the system.

Notice, that for small ${\bf k}$ we have $m^\prime \simeq m, \
q^\prime_{d(h)} \simeq q_{d(h)}$ and from dispersion relations
(\ref{2TRD-18}) and (\ref{2TRD-A7}) we obtain for degenerate
and hot fluids, respectively:
\[
\omega_{d(h)}^4 \ = \ \frac{k^2}{2m}\Big[2p_{d(h)}q_{d(h)}+r_{d(h)}^2
+ s_{d(h)}^2 \ \pm
\]
\begin{equation}
\pm \ (r_{d(h)}+s_{d(h)})\sqrt{(r_{d(h)}-s_{d(h)})^2+4p_{d(h)}q_{d(h)}}\ \Big]
\label{2TRD-24}
\end{equation}
and, consequently, the relations between fields read as:
\[
{\bf U}_{d(h)}=\Big[\frac{s_{d(h)}-r_{d(h)} }{2p_{d(h)}} \ \pm
\]
\begin{equation}
\qquad \pm \ \frac{1}{2p_{d(h)}}\,\sqrt{4p_{d(h)}q_{d(h)}
+(r_{d(h)}-s_{d(h)})^2} \ \Big]\ {\bf B} \ .
\label{2TRD-26}
\end{equation}

Thus, we see, that besides the ambient system-defining parameters,
the relations between macro-fields are strongly dependent also on
corresponding DR solution and range of ${\bf k}$. Below we show
how the Unified RD/D related phenomena influence the 
evolution dynamics of 2-Temperature relativistic e-i plasmas of astrophysical
objects.

\section{Unified Dynamo/Reverse Dynamo for %AWDs'
2-Temperature Relativistic e-i plasmas of astrophysical objects}

We performed the extensive numerical analysis study of the derived equations.
For an illustrative analysis we choose the characteristic case of WD with
degenerate effective mass $G_0=1.5$ (corresponding electron number density
$\sim 10^{30}\,cm^{-3}$). For an accreting hot contamination we present
results for two cases of $H_0=1.5$ and $H_0=10$ (corresponding electron
temperatures $2.2 \cdot 10^9$\,K and $1.5 \cdot 10^{10}$\,K,
respectively). As an example we considered two extreme cases
of following DB parameters:
(i) $a_d = - a_h \equiv a$ and (ii) $a_d=\frac{1}{a_h} \equiv a$; for all
cases $\alpha = 0.001$ since the detailed analysis showed
no significant dependence on $\alpha $ as long as it is \ $\leq 0.01$.

(i) For the case $a_d = -a_h \equiv a$ we have explored various
interesting cases among which we present below 2 simplest
examples for $k \to 0$ and characteristic limits for
Beltrami parameter $a$ so that for dispersion relations
we use equations (\ref{2TRD-24}) for degenerate (hot) fluids:

(i-1) For the range of $2.5 < a < 15$ we have ${\bf v}_{0h}\ll
{\bf b}_0 \ll {\bf v}_{0d} $ meaning that the ambient degenerate
flow is primarily kinetic while the ambient hot one is magnetically
dominant; the plots display the results for 2 different hot
fluid temperatures: $H_0 = 1.5$ (red and magenta) and $H_0=10$
(blue and green). In Fig.1 the Alfv\'en Mach numbers for
both fluids generated macro- and micro-scale fields are
displayed for the solutions of $ \omega_{d1}^4=\frac{k^2}{2m}
\Big(2p_dq_d+r_d^2+s_d^2+(r_d+s_d) \sqrt{(r_d-s_d)^2+4p_dq_d}\Big) $
(left column) and $\omega_{h1}^4=\frac{k^2}{2m}\Big(2p_hq_h+r^2_h+s^2_h
+ m(r_h+s_h)\sqrt{(r_h-s_h)^2+4p_hq_h}\Big) $ (right column),
while the results for the solutions of $ \omega_{d2}^4=\frac{k^2}{2m}
\Big(2p_dq_d+r_d^2+s_d^2-(r_d+s_d)\sqrt{(r_d-s_d)^2+4p_dq_d}\Big) $ (left column)
and $\omega_{h2}^4=\frac{k^2}{2m}\Big(2p_hq_h+r^2_h+s^2_h -
m(r_h+s_h)\sqrt{(r_h-s_h)^2+4p_hq_h}\Big) $ (right column)
are displayed in Fig.2. We observe, that dynamics is quite
different for these 2 cases - in the former case we get
for both-scale fields the Reverse Dynamo scenario with
Super-Alfv\'enic generated fields; specifically
hot fluid fluctuations are very fast (Mach numbers are within
$(1 - 10)$ for degenerate flow and $(1 - 5)\cdot 10^2$
for hot flow); for the latter case the scenario
is the one of Dynamo: degenerate fluid macro- as well as
micro-scale and hot fluid velocity fluctuations are no
more strong - we could call this scenario the Unified
Reverse Dynamo/Dynamo process for hot fraction when the
large scale hot-flow is Super-Alfv\'enic and the short-scale
one is sub-Alfv\'enic. The results are practically independent
of $H_0$ except for micro-scale hot fluid components (that
decrease with increasing $a$). The entire process is the
illustration of Unified Dynamo/Reverse Dynamo process for
a mixed system predicting the generation of strong magnetic
fields of both scales simultaneously to the weak flows/outflows
in the 2-temperature relativistic complex system. This may
explain the existence of strong magnetic fields in the
vicinity of accreting massive stars (see e.g.
\citep{Fossil,Binary-B} \& references therein).

%%%%%%%%%%%%%%%%%%% FIG.1 %%%%%%%%%%%%%%%%%%%%%%%

\begin{figure}
\includegraphics[width=0.38\textwidth]{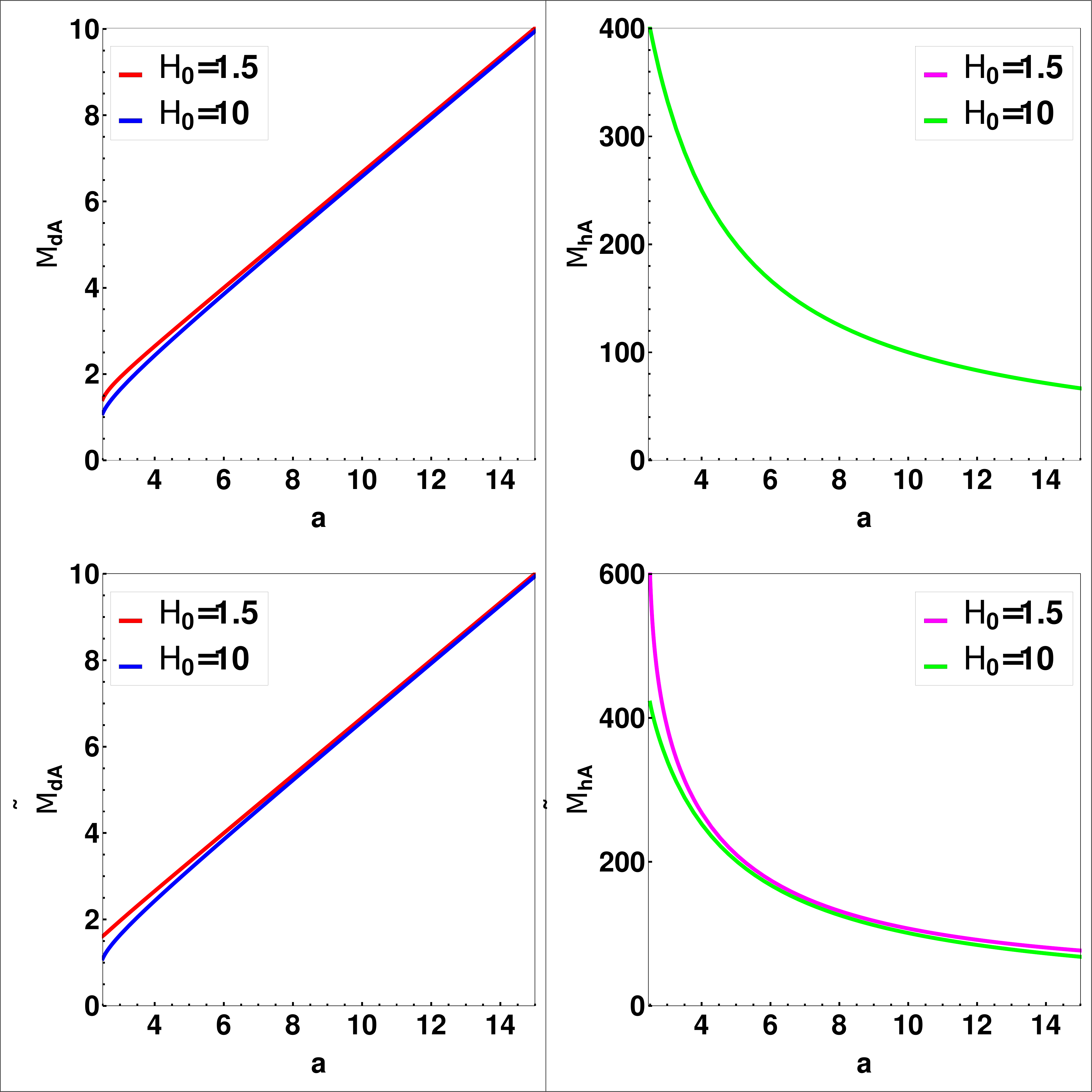}
\centering
\caption{Plots for Alfv\'en Mach numbers versus Beltrami parameter
$a_d=-a_h \equiv a$ (case (i-1)) for degenerate fluid (left column)
for the $\omega_{d1}$ solutions and for hot relativistic fluid
(right column) for the solutions of $\omega_{h1}$. Reverse Dynamo
for both fluids in both the macro- and micro-scale components -
hot flow is strongly Super-Alfv\'enic. Plots display the results
for 2 different hot fluid temperatures: $H_0=1.5$ (red and magenta)
and $H_0=10$ (blue and green). The results are practically independent
of $H_0$ except for micro-scale hot component Mach numbers that
decrease with increasing $a$.}
\label{fFig.1}
\end{figure}

%%%%%%%%%%%%%%%%%%%%%%%%%%%%%%%%%%%%%%%%

%%%%%%%%%%%%%%%%%%%%%%% FIG.2 %%%%%%%%%%%%%%%%%%%%%%%%%%%%%%%%%%%%

\begin{figure}
{\includegraphics[width=0.38\textwidth]{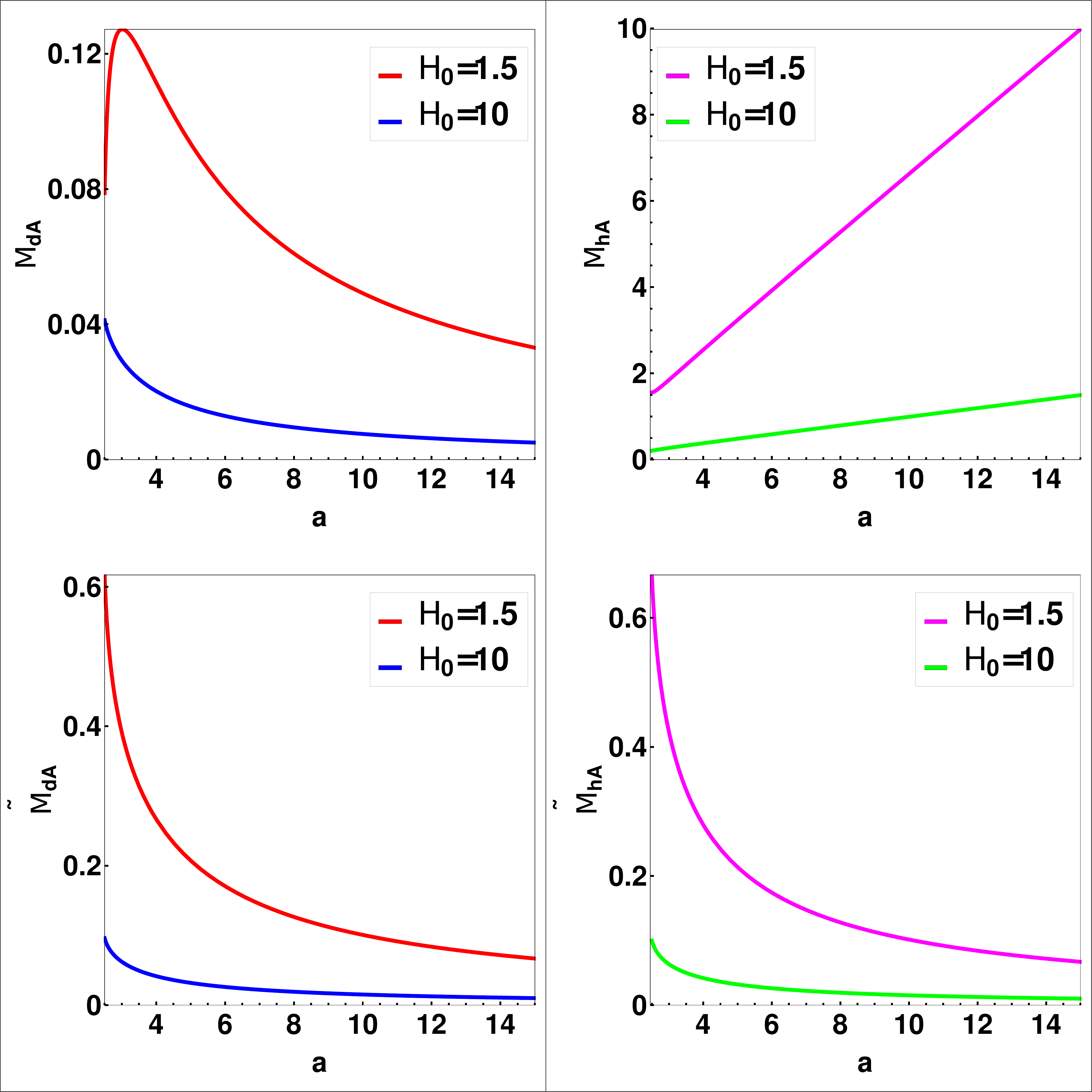}
\centering
\caption{Plots for Alfv\'en Mach numbers versus Beltrami parameter
$a_d=-a_h \equiv a$ (case (i-1)) for degenerate fluid (left column)
for the $\omega_{d2}$ solutions   and for hot  relativistic fluid
(right column) for the solutions of $\omega_{h2}$ - Dynamo for degenerate
and Unified D/RD for hot - Unified D/RD for entire mixed system.
Plots display the results for 2 different hot fluid temperatures:
$H_0=1.5$ (red and magenta) and $H_0=10$ (blue and green) that are
in the same range for each component.}
\label{fFig.2}}
\end{figure}

(i-2). For the range of $2.45\leq a < 2.5$ we have
${\bf b}_0 \gg {\bf v}_{0d}, \ {\bf v}_{0h} $
meaning that both degenerate fluid and relativistically
hot fluid-contamination are primarily magnetic; the plots
display the results for $H_0 = 1.5$ (red and magenta)
and $H_0=10$ (blue and green). In Fig.3 the Alfv\'en Mach
numbers for both fluids generated macro-
and micro-scale fields are displayed for the solutions
$ \omega_{d1}^4$ (left column) and $\omega_{h1}^4$
(right column), while the results for the solutions $ \omega_{d2}^4$
(left column) and $\omega_{h2}^4$ (right column) are displayed
in Fig.4. We again observe, that the dynamics is quite
different for these 2 cases - for the former case the
generated degenerate flows are are practically Alfv\'enic
in both scales while the hot flows are super-Apfv\'enic
(with Mach numbers within $(1 - 5)\cdot 10^2$ - straight
Reverse Dynamo scenario, independent of ambient temperature)
and for the latter case we have purely Dynamo process
in both fluids and both-scales - such scenario can explain
the generation of strong magnetic fields in the envelope
phase of star accretion, for instance cases of intermediate-field
WDs found in CVs and some HFMWDs are argued to be formed
through interface dynamo / disk dynamo processes
\citep{Ferrario-3,Disk-DB}. We remind the reader, that
such combined scenario is absent in pure degenerate e-i case
for the primarily magnetic ambient system, hence, such
possibility is entirely due to the hot contamination found
in accreting stars / binary system; the unified Dynamo/RD
scenario was found for degenerate e-i system only for
kinetic ambient system argued to explain the core-dynamo
\citep{ruderman,kissin} formation of magnetic \\
fields of HFMWDs \citep{RD_deg}.
Interestingly, existence of Super-Alfv\'enic large-scale
hot-flows for our composite system (with different solutions
of dispersion relations co-existing in both fluids) may
explain the formation of transient jets fed by short-scale
fluctuations of both fluids that follow Dynamo -
a well established path of Unified Dynamo/Reverse Dynamo.

%%%%%%%%%%%%%%%% FIG.3 %%%%%%%%%%%%%%%%%

\begin{figure}
\includegraphics[width=0.38\textwidth]{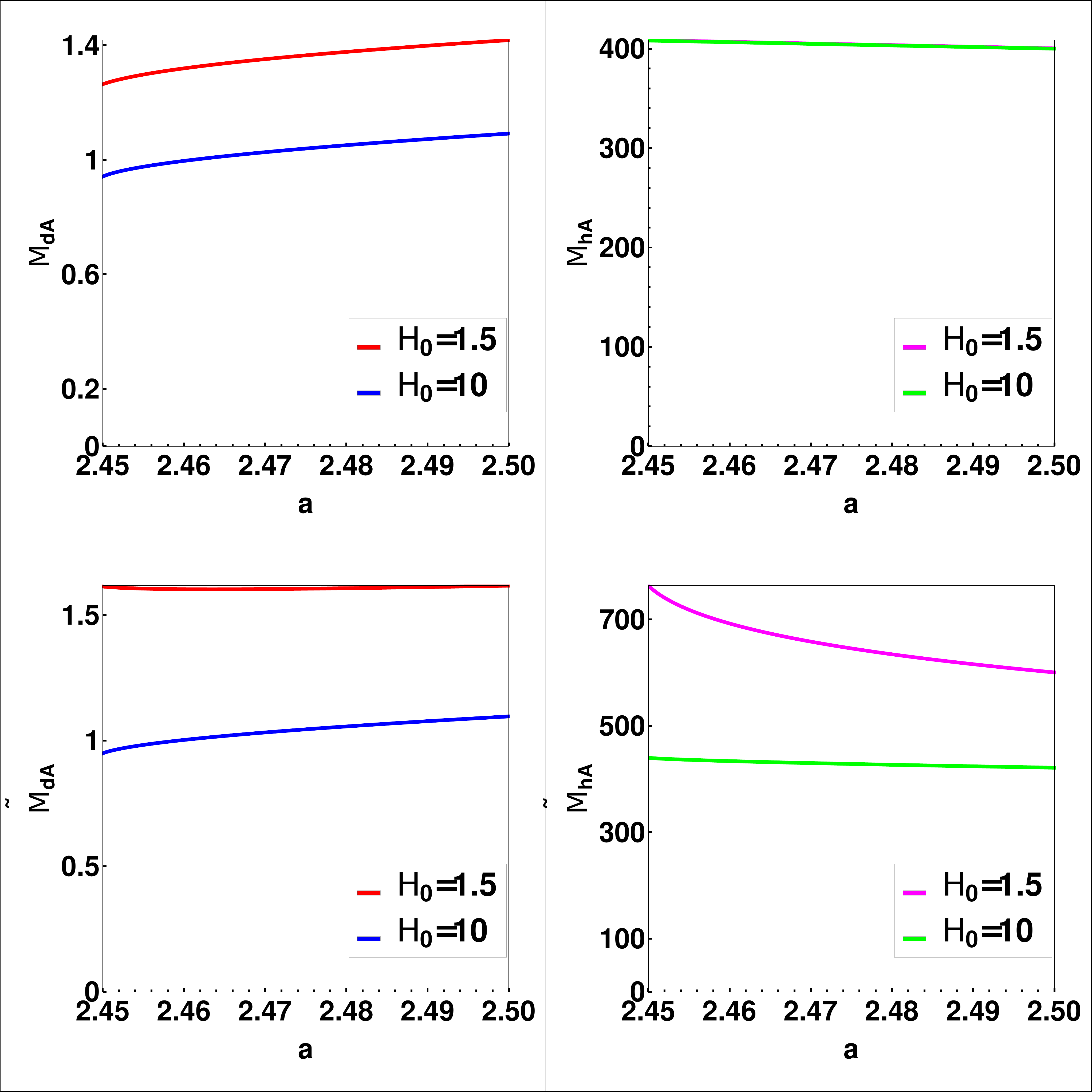}
\centering
\caption{Plots for Alfv\'en Mach numbers versus Beltrami parameter
$a_d=-a_h \equiv a $ (entire ambient system is magnetically dominant -
case (i-2))   for degenerate fluid (left column) for the $\omega_{d1}$
solutions and for hot   relativistic fluid (right column) for the
solutions of $\omega_{h1}$. Reverse Dynamo for a mixed system - degenerate
flow becomes practically Alfv\'enic while hot flow is strongly
Super-Alfv\'enic in both scales. The results for macro-scale
generated hot outflow coincide for different ambient temperatures.}
\label{fFig.3}
\end{figure}

%%%%%%%%%%%%%%%%%%%%% FIG.4 %%%%%%%%%%%%%%%%%%%%%%%%%%%%%%%%%%%%%%

\begin{figure}
{\includegraphics[width=0.38\textwidth]{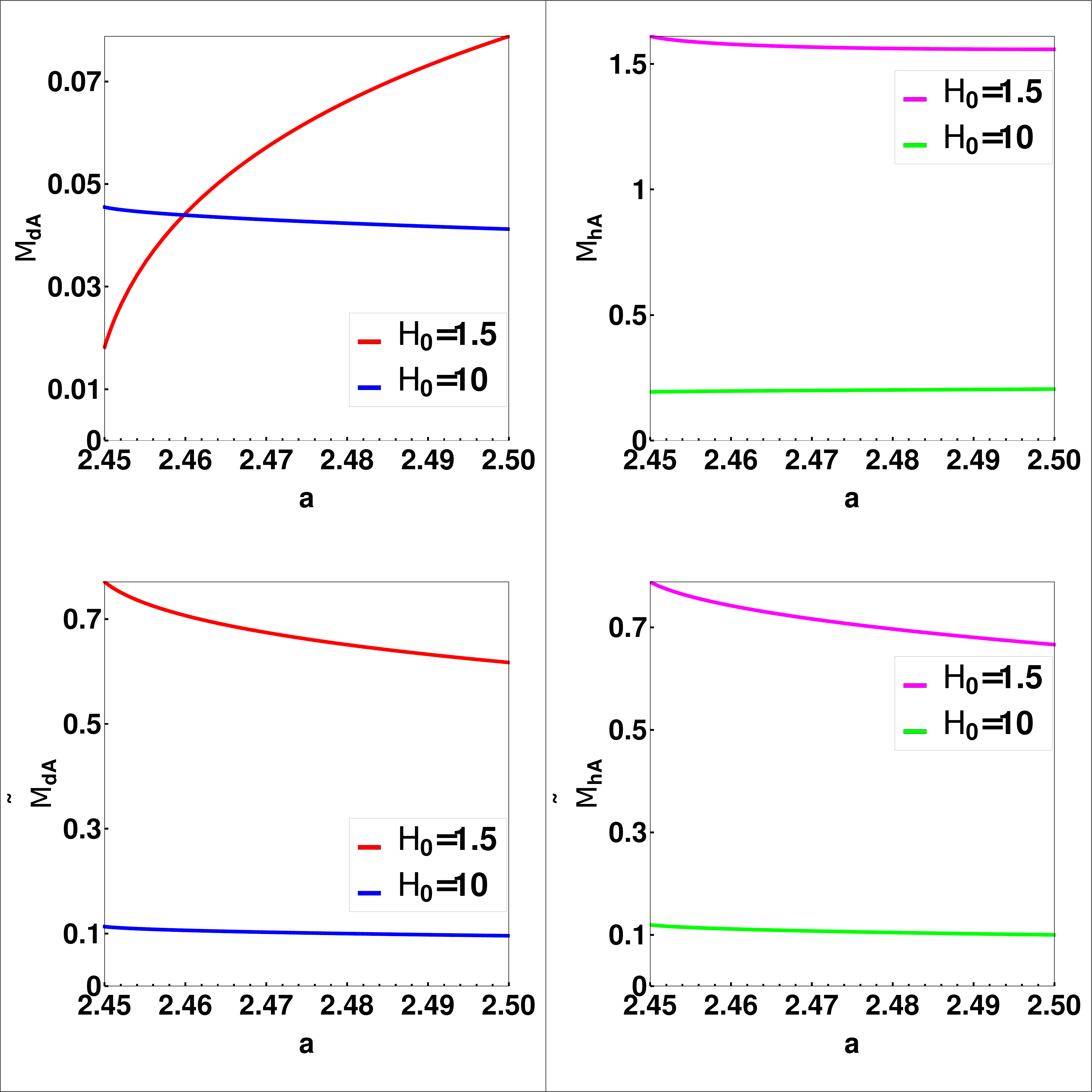}
\centering
\caption{Plots for Alfv\'en Mach numbers versus Beltrami parameter
$a_d=-a_h \equiv a$ (entire ambient system is magnetically dominant -
case (i-2))   for degenerate fluid (left column) for the $\omega_{d2}$
solutions and for hot   relativistic fluid (right column) for the
solutions of $\omega_{h2}$ - straight Dynamo for both fluids practically
in both scales (hot fluid large-scale component is nearly
Alfv\'enic).}
\label{fFig.4}}
\end{figure}

We have examined the dependence of final results on the hot relativistic
fraction ambient temperature $H_0$. As shown above for Reverse
Dynamo scenarios (Figures 1,3) the Mach numbers are practically
independent of $H_0$. In Fig.5 we present the maximal values of
Mach numbers for the 2nd roots of cases (i-1) (Fig.2) and (i-2)
(Fig.4); Beltrami parameter was chosen to be $a=10$ (blue, green -
 magnetically dominant degenerate fluid) for the
former case and $a=2.5$ (cyan, orange - magnetically dominant
entire ambient system) for the latter. The Maximal values
decrease with the increase of ambient temperature. Main result is the
same for the 2nd root: there is a possibility of strong magnetic field
generation while the WD envelope phase evolution
when accreting the hot astrophysical flow.

%%%%%%%%%%%%%%%%%%%%%%% FIG.5 %%%%%%%%%%%%%%%%%%%%%%%%%%

\begin{figure}
{\includegraphics[width=0.38\textwidth]{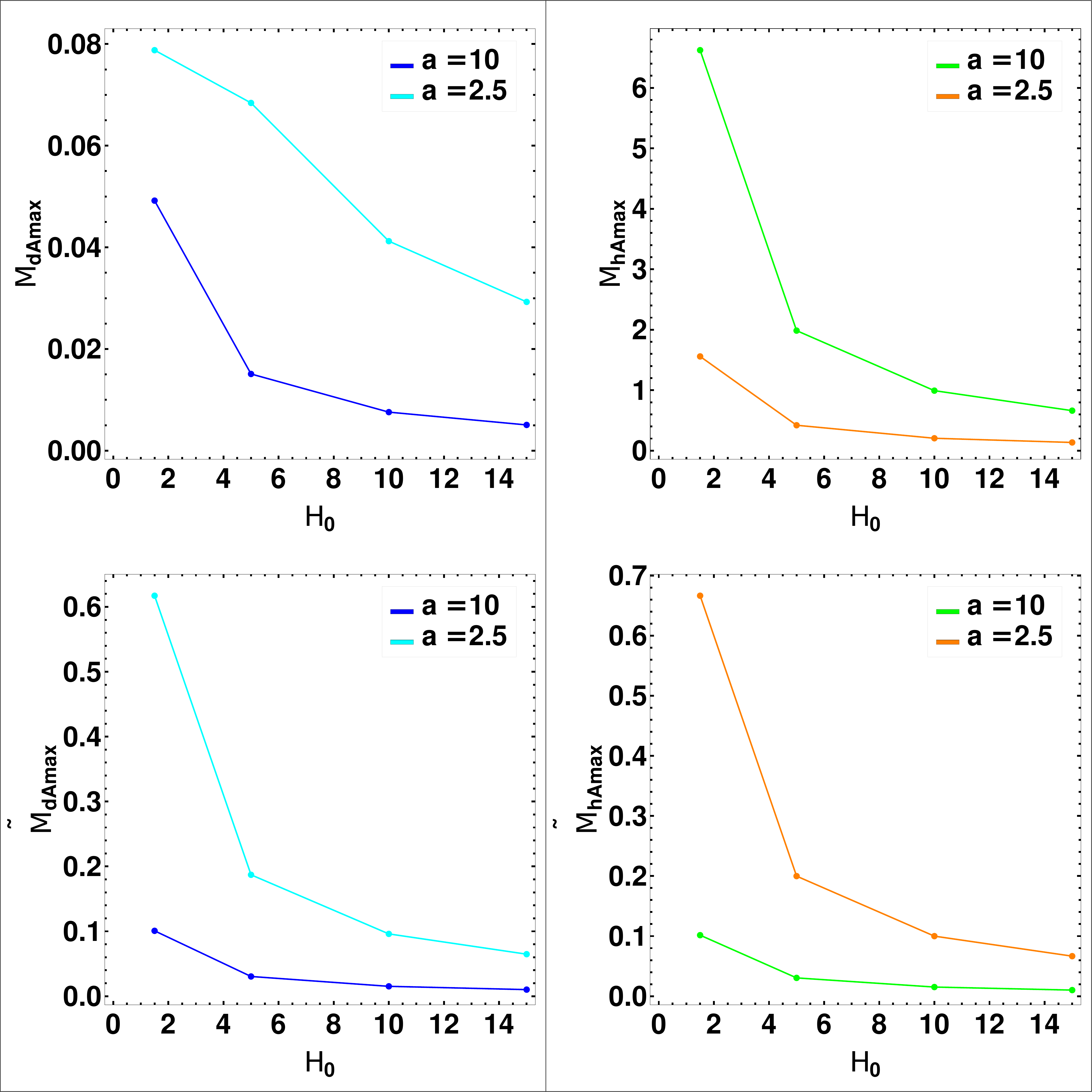}
\centering
\caption{Maximal values of Alfv\'en Mach numbers versus relativistic
hot fraction ambient temperature $H_0$ for the cases plotted in:
Fig.2 (cyan and blue) and Fig.4 (orange and green) for $a_d=-a_h\equiv a$;
results are displayed for 2 limiting ambient system cases: $a = 2.5$
(cyan, orange - magnetically dominant entire system) and $a=10$
(blue, green - magnetically dominant degenerate fluid)
for degenerate fluid (left column) and for hot relativistic fluid
(right column)  -   the bigger the $H_0$ the smaller is the maximal
value for all scales in both fluids.}
\label{fFig.5}}
\end{figure}

%\bigskip

\begin{figure}
\begin{center}
\includegraphics[scale=0.2,angle=0]{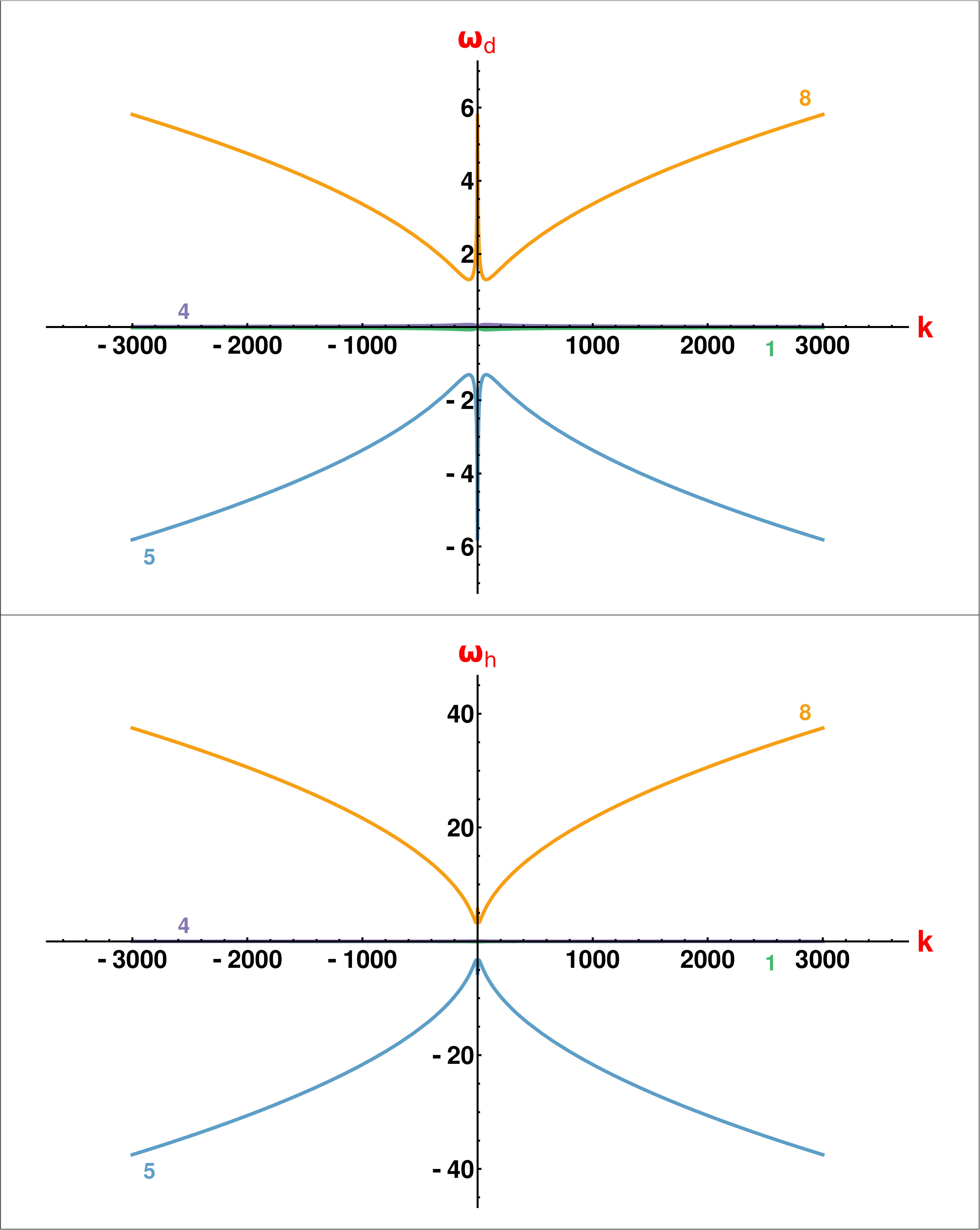}
\caption{Solution of dispersion relations (\ref{2TRD-18})
(degenerate electron flow - top) and (\ref{2TRD-A7})
(hot electron flow - bottom) for case (ii-1): $a_d =
\frac{1}{a_h} \equiv a = 10$ - bulk ambient degenerate (hot fraction)
system is kinetically (magnetically) dominant; $G_0=1.5, \ H_0=10$
(picture is similar for $H_0=1.5$); 4  real roots at $q\prime
\simeq q$ are displayed by different color. }
\label{ffig.6.}
\end{center}
\end{figure}

(ii) For case $a_d = 1/a_h \equiv a$ we have explored various
interesting cases among which we present below 2 simplest
important examples for $k \neq 0$ and characteristic limits
for Beltrami parameter $a$ so that for DRs we use general equations
(\ref{2TRD-18}) for degenerate and (\ref{2TRD-A7}) for hot flows,
respectively; we solved them numerically and found corresponding
inverse micro-scales for all scenarios. 2 characteristic examples
of such solutions are displayed in Fig.\ref{ffig.6.} and Fig.\ref{ffig.8.}
for specific ambient system conditions, different colors correspond
to 4 real roots. Extensive study for different parameters showed
that dispersion picture does not change much - we have always
four real roots with similar ${\bf k}$-dependence.

%%%%%%%%%%%%%%%%%  FIG.6   %%%%%%%%%%%%%%%%%%%%%

\begin{figure}
\begin{center}
\includegraphics[scale=0.4,angle=0]{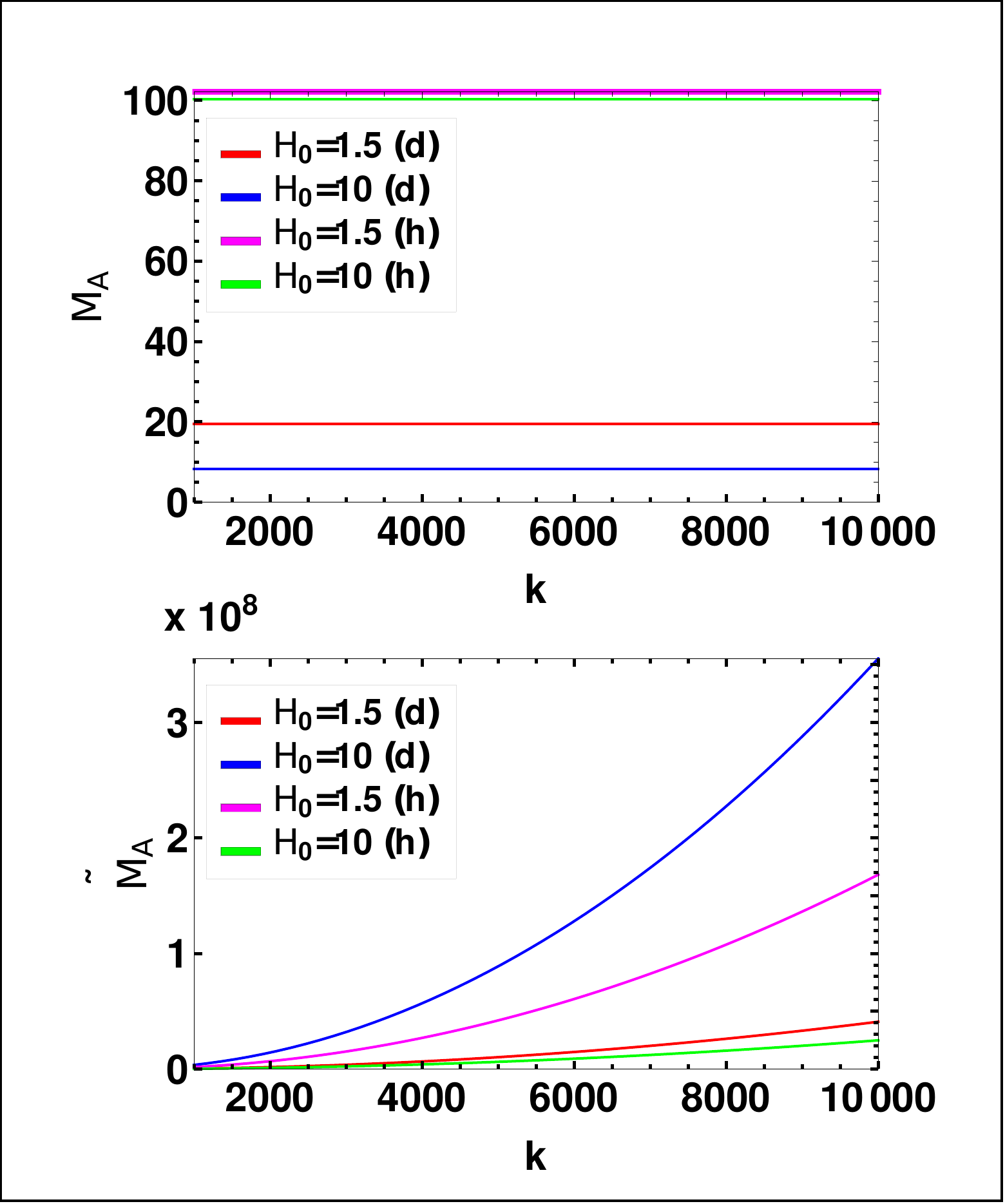}
\caption{Alfv\'en Mach numbers versus $k$ for:
macro-scale vector-fields $M_A$ (top) and micro-scale vector-fields
${\tilde{M}_A}$ (bottom), respectively
for generated velocity and magnetic fluctuations
for the root 1 of Fig.6;
$a=10$. Bigger the $k$ bigger is $\tilde{M}_A$
while macro-scale velocity fields are practically independent of $k$
for the same $H_0$; generated macro(micro) flows are strongly super-Alfv\'enic
for both fluids (red, blue - degenerate, magenta, green - hot)
- Unified Reverse Dynamo/Dynamo scenario.}
\label{ffig.7.}
\end{center}
\end{figure}

\begin{figure}
\begin{center}
\includegraphics[scale=0.2,angle=0]{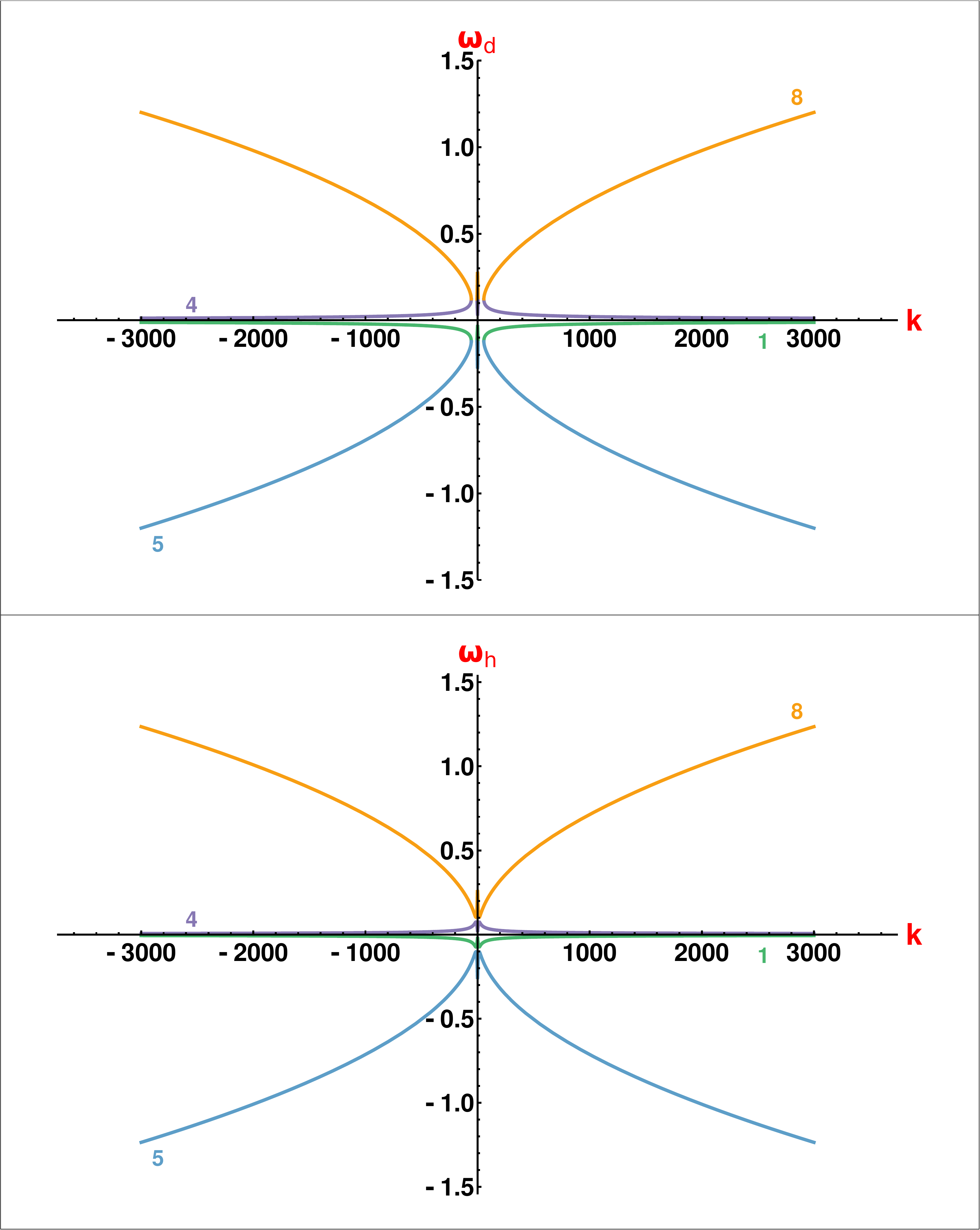}
\caption{Solution of dispersion relations (\ref{2TRD-18})
(degenerate electron flow - top) and (\ref{2TRD-A7})
(relativistically hot flow - bottom) for the (ii-2) case:
$a_d = \frac{1}{a_h} \equiv a = 2.45$ - entire ambient system
is magnetically dominant; $G_0=1.5, \ H_0=10$ (solutions
show similar picture for $H_0=1.5$); 4  different real roots
at $q\prime \simeq q $  are displayed by different color. }
\label{ffig.8.}
\end{center}
\end{figure}

(ii-1) For $a\sim10 $ - ambient energy is kinetically dominant
only for degenerate flows ($ {\bf v}_{0h} \ll {\bf b}_0 \ll {\bf v}_{0d} $).
In Fig.\ref{ffig.7.} Alfv\'{e}n Mach numbers are plotted for
roots 1 of Fig.\ref{ffig.6.}, where we observe generation of
Super- Alfv\'{e}nic velocity fields of hot flow
for two different temperatures  $H_0=1.5$ and $H_0=10$;
for roots 8 of dispersion relations (not displayed here since
more representative result for roots 8 is displayed in Fig. 9
for (i-2) case) of the same Fig.6 we found that much stronger
Super-Alfv\'{e}nic macro- and micro-scale hot flows ($>10^6$)
are generated for both temperature electrons.
This is a perfect scenario for Unified Reverse Dynamo/Dynamo
phenomenon both in macro- and micro-scales since the
large-scale magnetic fields are generated simultaneously
due to magneto-fluid coupling. Notice, that for all roots
of DR the generated hot flows are stronger (2 orders stronger)
than degenerate flows (in both scales).

%%%%%%%%%%%%%%%%%%  Fig.7  %%%%%%%%%%%%%%%%%%%%%%%%%%%%%%%

(ii-2) For $a\sim 2.45 $ - whole ambient energy is mostly
magnetic ($ {\bf b}_0 \gg {\bf v}_{0d}, \ {\bf v}_{0h} $).
In Fig.9 Alfv\'{e}n Mach numbers are plotted for roots 8 of
Fig.8, where we observe generation of very strong Super-Alfv\'{e}nic
velocity fields of hot flow for two different temperatures
$H_0=1.5$ and $H_0=10$ - both macro- and micro-scale fast hot
flows ($\gtrsim 10^8$) are generated for both temperatures.
We also observe, that again we have very fast flows for roots 8
while for roots 1 (not displayed here), interestingly, 
numerical analysis for the illustration of the qualitative study
show, that $M_A$ is practically independent of $k$ (like for
case displayed in Fig.7 of (i-1) parameter range).
These results could be easily explained by the dispersion relation
displayed in Fig.8 - we see that for roots 1 (green) there
is no significant change for big $k$-s, but for roots 8
(yellow) picture is very different.

%%%%%%%%%%%%%%%%%  FIG.8   %%%%%%%%%%%%%%%%%%%%%

%%%%%%%%%%%%%%%%%%%%%%%%  FIG.9 %%%%%%%%%%%%%%%%%%%%%%%%%%%%%%%%%%

\begin{figure}
\begin{center}
\includegraphics[scale=0.38,angle=0]{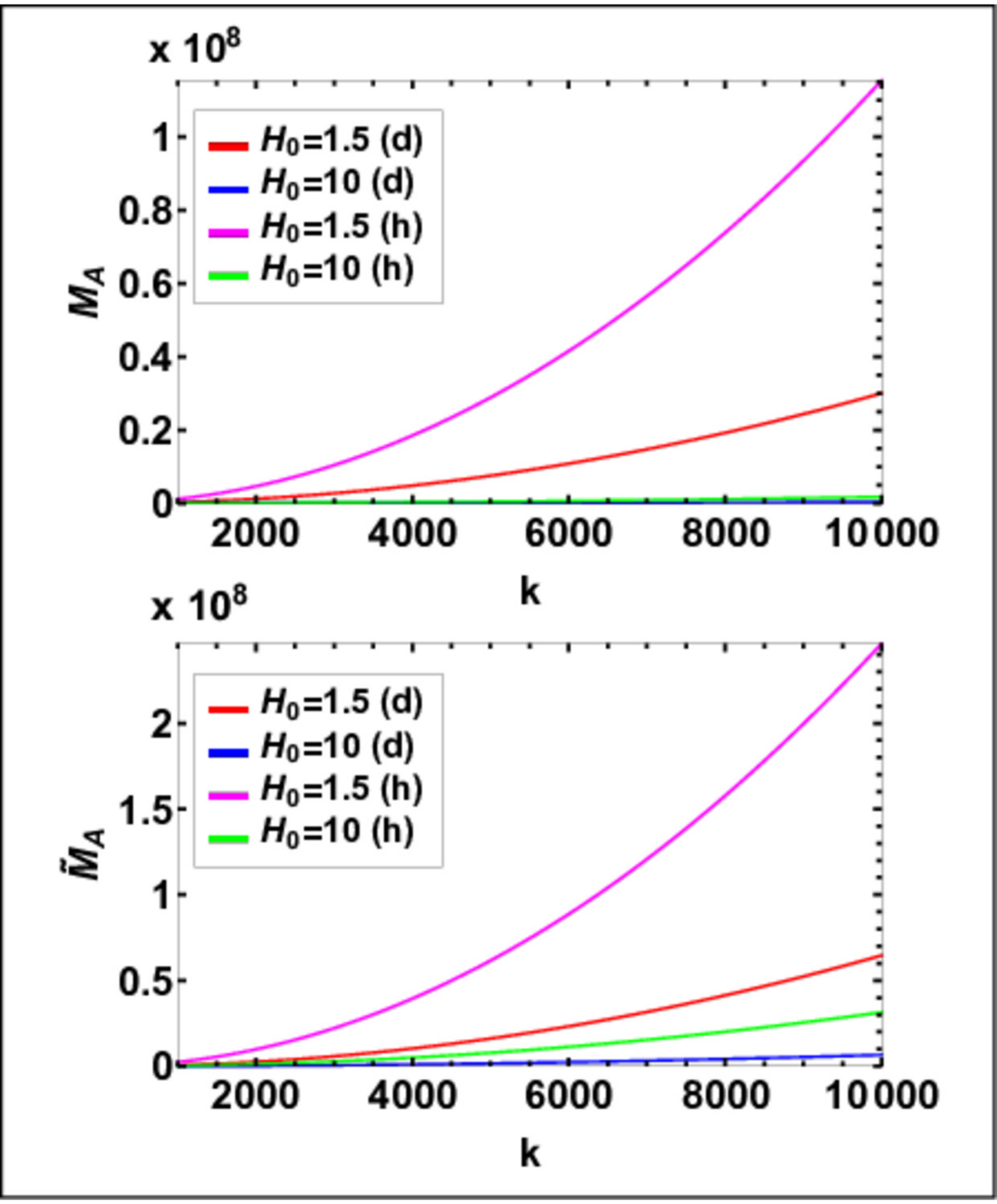}
\caption{Alfv\'en Mach numbers versus $k$ for:
macro-scale vector-fields $M_A$ (top) and micro-scale vector-fields
${\tilde{M}_A}$ (bottom), respectively  for generated
velocity and magnetic fluctuations for the roots 8 of Fig.8;
$a=2.45$. Bigger the $k$ bigger are both fluids' $M_A$ and $\tilde{M}_A$;
both macro- and micro-scale velocity fields are strongly
super-Alfv\'enic for both fluids ($\gtrsim 10^8$) - manifestation of
perfect Unified Reverse Dynamo/dynamo scenario in both scales
(the hot flow $M_A$-s and $\tilde{M}_A$-s are multiplied by
$10^{-2}$ for best illustration).}
\label{Fig.9.}
\end{center}
\end{figure}

We have examined the dependence of final results on the ambient
relativistic hot electron fluid fraction temperature $H_0$.
Results for roots 8 are presented in Fig.10 showing that
maximal macro-scale Alfv\'en Mach number $M_{Amax}$ increases
when decreasing $H_0 $ and can reach the values $\sim 10^8$ for
degenerate electron fluid ($\sim 10^{10}$ for hot electron fraction)
for $H_0=1.5 $ at $k \lesssim 10^4$.  This is a perfect manifestation
of Unified Reverse Dynamo / Dynamo scenario in both scales of
both fluids.

\bigskip

Thus, analysis showed that the simultaneous dynamical generation
of strong macro-scale magnetic field and fast flows/outflous
is guaranteed in 2-temperature relativistic system consisting of
degenerate e-i fluid and relativistically hot fluid contamination due to
magneto-fluid coupling. Depending on the range of Beltrami scales
(helicities) of degenerate and hot fluids ($a_d$ and $a_h$) as well as density
degeneracy level for former ($G_0$) and the temperature of latter ($H_0$)
the ratio between macro-scale velocity and magnetic fields
may become different for each fluid and for such a mixed system
evolution scenarios maybe of any character; similar conclusions
can be drawn for micro-scale fields. There are various scenarios
for WD evolution for any type mixture of ambient bulk fluid
and contamination; e.g. for specific parameter ranges two possibilities are:
1) Unified D/RD for one of the roots of DRs resulting in generation
of strong magnetic fields while accretion and star formation in binary
systems \citep{Garcia,MergingBinary}.
2) Unified RD/D for some other roots of DRs resulting
in generation of fast (with Mach numbers $> 10^6$) flows/outflows
\citep{beskin,Garcia2}; $M_{Amax}$ is higher for $a_d=(1/a_h)$
than for $a_d=-a_h$.
At the end, there can be any mixture of macro- and micro-
fields in both fluids over the time but as long as the ambient
hot flow fraction is magnetically dominant independently
from degeneracy level and energetic character of bulk degenerate
ambient fluid a formation of super-fast hot outflows (with
$M_{hAmax}>10^8>M_{dAmax}$) is guaranteed while WD evolution.
Fraction parameter $\alpha \ll 1$ value doesn't play significant role.
Note, that for the global dynamics of multi-component and
multi-scale plasmas accreting into WDs, in addition to the discussed
parameters the corotating/counterrotating orbits of plasmas, the
mass ratio between the binaries, accretion rate, density/temperature
profiles, etc. will play significant role [see e.g. \citep{SPH,liu,3Dsim}
for the extensive SPH simulations of the compact objects/binary
accreting system global dynamics].

%%%%%%%%%%%%%%%%%%%% FIG.10 %%%%%%%%%%%%%%%%%%%%%%%%%%%%

\begin{figure}
{\includegraphics[width=0.38\textwidth]{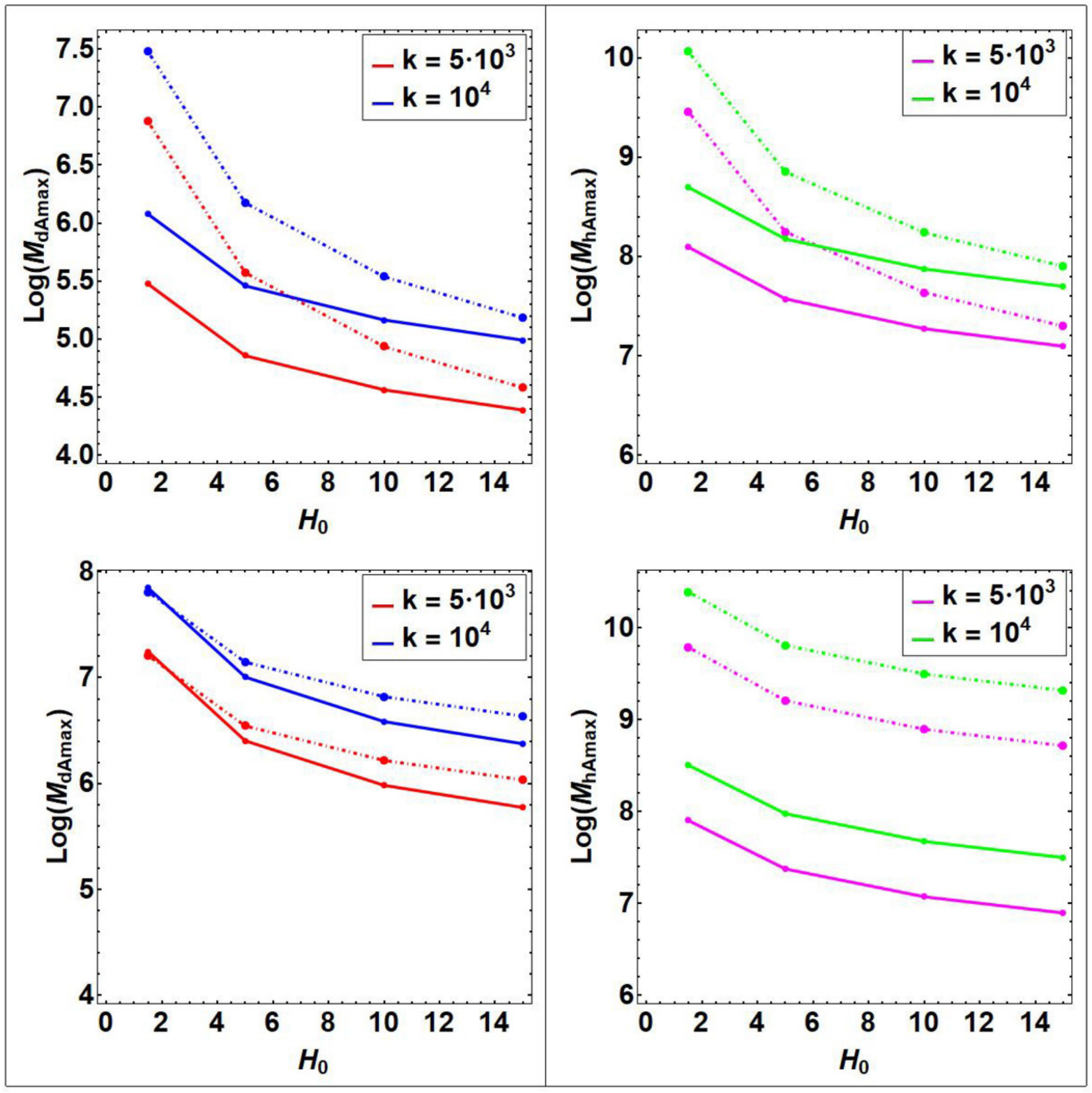}}
\centering
  \caption{Maximal values of Alfv\'en Mach Numbers versus $H_0$
for: macro-scale vector-fields $M_{Amax}$ (top) and micro-scale vector-fields
${\tilde{M}_{Amax}}$ (bottom), respectively for the roots 8 of Fig.6
(solid lines, $a=10 $) and roots 8 of Fig.8 (dotted lines, $a=2.5 $
- magnetically dominated ambient system). Smaller the $H_0$ bigger
is the $M_{Amax}$ that can reach values $\gtrsim 10^6$  at small $k$ .
Red, blue (magenta, green) colors correspond to:
$k=5\cdot 10^3 ; \ 10^4 $, for degenerate (hot) fluid, respectively.
Smaller the $k$ bigger is $M_{Amax}$ for the same $H_0$. Macro-scale
Mach numbers for hot fluid are 2 orders higher than those for degenerate.}
\label{fFig.10.}
\end{figure}

\section{Conclusions}

From an analysis of relativistic 2-Temperature electron-ion plasma
(bulk degenerate e-i fluid with a small fraction of relativistically
hot e-i, e.g. like in White Dwarfs accreting
hot astrophysical flow / Binary systems)
we have extracted the acceleration / generation / amplification of
the macro-scale flow/outflow and magnetic field due to Unified
Reverse Dynamo/Dynamo Mechanism in the composite astrophysical
systems with initial turbulent (micro-scale) magnetic/velocity
fields. This process is simultaneous with and complementary to
the micro-scale unified D/RD or RD/D dynamics like in two-fluid
cases - consequence of magneto-fluid coupling. Generation
of macro-scale fast flows (in both degenerate and hot fluids)
and strong magnetic fields are simultaneous:
the greater the macro-scale magnetic field (generated locally)
the greater becomes the macro-scale velocity field (generated
locally). Principle results are following:
\begin{itemize}
\item {Both bulk degenerate fluid and hot contamination
undergo the Unified RD/D dynamics; final picture is
complex leading to acceleration/amplification
of both fluid velocity/magnetic fields.}
\item {Bulk as well and fraction component flow/outflow acceleration
and magnetic field amplification due to the Unified RD/D
is directly proportional to initial turbulent kinetic or/and magnetic
energy in 2-temperature e-i relativistic astrophysical plasma;
such flows/magnetic fields are fed by either initial turbulent state:
fully kinetically or magnetically dominated ambient system or the different
combinations for different fluids. The generated/accelarated outflows
are extremely strong when both fluids are magnetically dominant; hot
outflows are several orders stronger than the degenerate ones.}
\item{Along with degeneracy level  ($G_0(n)$)
of bulk system and temperature of a hot fraction ($H_0$)
the scenarios are different for different Beltrami parameters
($a_d , \ a_h$) but there always exists such a real $\omega_d (\omega_h) $
(solution of corresponding fluid DR)
for which the generation of fast macro-scale locally
Super-Alfv\'enic outflow in one of the fluids or in both fluids
or/and the strong macro-scale magnetic field is guaranteed. }
\item{Formation process is less sensitive to hot fluid fraction
parameter $\alpha \ll 1$ but more sensitive to its temperature and
the corresponding solution of DR - for small $k$ the real roots of
DR define the processes of either straight D or RD; for big $k$ the
generation of strong macro-scale fast, locally Super-Alfv\'{e}nic
flow is guaranteed; for the same degeneracy state of the bulk system,
fraction ratio and the magneto-fluid coupling ($G_0$, $\alpha $ and
$a_d , \ a_h$), the smaller is fraction temperature ($H_0$)
larger is the corresponding fluid macro(micro)-scale Alfv\'en
Mach number and, hence, generated flow/outflow (magnetic field)
will have a dispersion with Velocity (Magnetic) field distribution
in (${\bf r}, t$) - observations show that large-scale astrophysical
flows as well as the magnetic fields are very complex with a characteristic
evolution in time and space. The generated macro-scale hot
outflow is stronger than the degenerate one for all solutions.}
\item{For some regimes of ambient system parameters (magnetically and/or
kinetically dominant mixed ambient system case), the
growing accelerated flows in both fluids are sub-Alfv\'enic - scenario
leading to strong magnetic field formation that grows as well -
this could explain the formation of macro-scale magnetic fields
while the envelope phase of star accretion / binary systems / WD evolution. }
\item{Specifically interesting finding comes to fully magnetically
dominant ambient system (for both fluids) when major part of its
energy is transformed into the extremely fast super-Alfv\'enic
macro-scale outflow energy due to magneto-fluid coupling for all
the real roots of dispersion relation; a weak macro-scale magnetic
field is generated along with it - this result is entirely due
to the 2-temperature character of ambient composite relativistic
system; for realistic physical parameters the resulting accelerated
generated locally super-Alfv\'enic flows are extremely fast with
Alfv\'en Mach number $> 10^6$ as observed in a variety of relativistic
astrophysical outflows.}
\end{itemize}

Thus, an intrinsic tendency of flow acceleration / magnetic field
amplification due to magneto-fluid coupling in multi-temperature
multi-component systems guarantees the simultaneous formation of
macro-scale fast flows/outflows and strong magnetic fields through
the Unified Reverse Dynamo/Dynamo mechanism - one possible
root to understand the evolution of accreting astrophysical
objects / binaries of different nature  -
one of the accessible sources to be considered together with
other additional formation mechanisms for macro-scale fast
flows/outflows as well as the macro-scale magnetic fields
(e.g. energy transformations due to catastrophe or waves in which
the  gravity, density/temperature inhomogeneities, rotation
play important role).

\section{Acknowledgements}

K.K.-s work was partially supported by World Federation
of Scientists National Scholarship Programme Geneva, 2020.
This work was partially supported by Shota Rustaveli
Georgian National Foundation Grant Project
No. FR17-391.

\section{The Data Availability Statement}

The data to support the outcomes of the work are openly available
by the authors upon the reasonable request sent to them directly.

\appendix

\section{Appendix - Derivation of Unified Dynamo/RD relations
for relativistically hot fluid fraction}

%\bigskip

Following the similar procedure as used for degenerate fluid
we derive the equations for hot relativistic
fluid fraction velocity field (below ${\bf Q}_h$ and ${\bf S}_h$
are functions of ${\bf U}_h$ and ${\bf B}$, as well as
plasma-system parameters):
\begin{equation}
m \frac{\partial \tilde{\bf b}}{\partial t} \ +
\ m_1\frac{\partial }{\partial t}\nabla \times \nabla \times {\tilde{\bf b}}\ =
\ ({\bf Q}_h \cdot \nabla){\bf b}_0 \ ,
%\label{2TRD-A1}
%\end{equation}
%\begin{equation}
\qquad \qquad
m \frac{\partial {\tilde{\bf v}}_{h}}{\partial t} \ +
\ q_1\frac{\partial }{\partial t}\nabla \times \tilde{\bf b} \ =
\ ({\bf S}_h \cdot \nabla){\bf b}_0 \ ,
\label{2TRD-A2}
\end{equation}
%and
%\[
\begin{equation}
{\rm and} \qquad \frac{G_0}{\alpha}\bigg(G_0+\frac{H_0}{\alpha}\bigg){\nabla\times\ddot{\textbf B}} \ ,
+\bigg(G_0+\frac{H_0}{\alpha}\bigg)^2{\ddot{\textbf U}}_h
%\]
%\begin{equation}
\ = \ q_h\nabla\times\textbf B + s_h\nabla\times\textbf U_h \ ,
\label{2TRD-A3}
\end{equation}
%\[
\begin{equation}
\qquad \qquad \frac{G_0H_0}{\alpha}\bigg(G_0+\frac{H_0}{\alpha}\bigg){\nabla\times
\nabla\times\ddot{\bf B}}+\bigg(G_0+\frac{H_0}{\alpha}\bigg)^2{\ddot{\textbf B}} \ = \
%\]
%\begin{equation}
%=
r_h\nabla\times {\bf B} + p_h\nabla \times {\bf U}_h \ ,
\label{2TRD-A4}
\end{equation}
\begin{equation}
m_1{\nabla\times\nabla\times\ddot{\textbf B}}+m{\ddot{\textbf B}}
= r_h\nabla\times\textbf B + p_h\nabla\times\textbf U_h \ ,
%\label{2TRD-A5}
%\end{equation}
%\begin{equation}
\qquad \qquad q_{1h}{\nabla\times\ddot{\textbf B}}+m{\ddot{\textbf U}}_h
= q_h\nabla\times\textbf H+s_h\nabla\times\textbf U_h \ .
\label{2TRD-A6}
\end{equation}
This time Dispersion relation reads as follows:
\begin{equation}
\qquad \omega_h^8 m^{\prime 2} m^2-\omega_h^4k^2\Big(2m^\prime mp_hq^\prime_h
+ m^2r^2_h+m^{\prime 2}s^2_h\Big)+(p_hq^\prime_h-r_hs_h)^2k^4=0
\label{2TRD-A7}
\end{equation}
%and
\begin{equation}
\qquad \qquad {\rm and} \qquad \qquad \textbf U_h \ = \ \frac{q^\prime_h\Big(\omega^4 m^\prime m-(p_hq^\prime_h-r_hs_h)k^2\Big)}
{\omega^4 m^2r_h+s_h(p_hq^\prime_h-r_hs_h)k^2} \ {\bf B} \ ,
\label{2TRD-A8}
\end{equation}
%where
\[
\qquad \qquad {\rm where} \qquad  q_{1h}=\frac{G_0}{\alpha}\bigg(G_0+\frac{H_0}{\alpha}\bigg) \equiv \frac{G_0}{H_0}q_1  \ ,
\ \ \ q^\prime_h=(q_h+q_{1h}\omega^2) \ ,
%\]
\qquad {\rm as \ before} \ m^\prime=(m+m_1k^2) \ {\rm and}
%$m^\prime=(m+m_1k^2)$ and
\]
\[
r_h=\alpha\frac{\lambda b^2_0}{3}\bigg[-\frac{(H_0-G_0) }{\alpha^2}
- \frac{G_0H_0\lambda(\chi_h+\lambda) }{\alpha^2} \ +
%\]
%\[
%\qquad +
\ \alpha\bigg(\frac{(H_0-G_0)\chi_h}{\alpha}+\frac{H_0}{\alpha^2}\lambda\bigg)^2\bigg] \ ,
\]
\[
p_h=-\alpha\frac{\lambda b^2_0}{3}\bigg(-\frac{(H_0-G_0)}{\alpha}-\frac{G_0H_0\lambda(\chi_h+\lambda)}
{\alpha}\bigg)\cdot
\]
\[
\qquad \cdot \bigg[-\frac{(H_0-G_0)}{\alpha\lambda}-\frac{G_0H_0(\chi_h+\lambda)}{\alpha}
+\lambda G_0 \ -
%\]
%\[
%\qquad -
\ \alpha \chi_h\bigg(\frac{H_0}{\alpha^2}-G_0\bigg)-(H_0-G_0)\chi_h+\frac{H_0}{\alpha}\lambda\bigg] \ ,
\]
\[
q_h=-\alpha\frac{\lambda b^2_0}{6}\bigg[\bigg(\frac{(H_0-G_0)\chi_h}{\alpha}+\frac{H_0}{\alpha^2}\lambda\bigg)
\xi_{1h}-\frac{\lambda G_0}{\alpha^2} \ +
%\]
%\[
%\qquad +
\ \frac{\chi_h}{\alpha}\bigg(\frac{H_0}{\alpha^2}-G_0\bigg)\bigg] \ ,
\]
%\[
\begin{equation}
s_h=-\alpha\frac{\lambda b^2_0}{6}\bigg[\bigg(\lambda \frac{G_0}{\alpha}
- \chi_h\bigg(\frac{H_0}{\alpha^2}-G_0\bigg)\bigg) %\cdot
%\]
%\[
%\qquad \cdot
\bigg(\frac{1}{\alpha\lambda}+\lambda G_0
- \alpha \chi_h\bigg(\frac{H_0}{\alpha^2}-G_0\bigg)\bigg) \ -
%\]
%\begin{equation}
%\qquad - \
\bigg(\frac{(H_0-G_0)}{\alpha}
+\frac{G_0H_0\lambda(\chi_h+\lambda)}{\alpha}\bigg)\xi_{1h}\bigg]
 \label{2TRD-A9}
\end{equation}
with
\begin{equation}
\xi_{1h}=\bigg(\frac{(1 + 2 G_0 \lambda^2 - 2 G_0^2 \eta^{\prime} \lambda^2)}{\alpha} -
 2 G_0 ( G_0 \eta^{\prime} - 1) \lambda \chi_h + \frac{
 G_0 \eta^{\prime} ( H_0 \lambda ( \chi_h-\lambda )-1)}{\alpha^2}\bigg) \ ,
%\label{2TRD-A10}
%\end{equation}
%\begin{equation}
\qquad \qquad \eta^{\prime}=\frac{G_0}{\alpha q_{1h}} \ .
\label{2TRD-A11}
\end{equation}

%\bigskip

For small $\textbf k$ leading to  $m^\prime=m, \  q^\prime=q$ and for dispersion we have:
\begin{equation}
\qquad \omega_h^8 m^4-\omega_h^4k^2m^2\Big(2p_hq_h+r^2_h+s^2_h\Big)
+(p_hq_h-r_hs_h)^2k^4=0 \ ,
\label{2TRD-A12}
\end{equation}
%with
\begin{equation}
\qquad {\rm with} \qquad \omega_{h1(2)}^4=\frac{k^2}{2m}\Big(2p_hq_h+r^2_h+s^2_h\pm(r_h+s_h)
\sqrt{(r_h-s_h)^2+4p_hq_h}\Big)
\label{2TRD-A13}
\end{equation}
and consequently the relation between fields is:
\begin{equation}
\qquad \qquad {\bf U}_h=\frac{s_h-r_h\pm\sqrt{4p_hq_h+(r_h-s_h)^2}}{2p_h}\ {\bf B} \ .
\label{2TRD-A14}
\end{equation}

\end{document}